\documentclass[aps,prl,twocolumn,superscriptaddress,preprintnumbers,nofootinbib]{revtex4-1}
\usepackage{amsmath,amssymb,graphicx,natbib,bm,enumerate,hyperref,color,breakurl}

\newcommand{\GeV}{\ensuremath{\mathrm{GeV}}}

\begin{document} 

\preprint{ANL-HEP-PR-13-45}
\preprint{EFI-13-21}
\preprint{KIAS-P13047}

\title{Very Light Charginos and Higgs Decays}

\author{Brian Batell}
\affiliation{Enrico Fermi Institute $\&$ Department of Physics, University of Chicago, Chicago, IL
60637, USA}

\author{Sunghoon Jung}
\affiliation{Enrico Fermi Institute $\&$ Department of Physics, University of Chicago, Chicago, IL
60637, USA}
\affiliation{School of Physics, Korea Institute for Advanced Study, Seoul 130-722, Korea}

\author{Carlos E. M. Wagner}
\affiliation{Enrico Fermi Institute $\&$ Department of Physics, University of Chicago, Chicago, IL
60637, USA}
\affiliation{HEP Division, Argonne National Laboratory, 
Argonne, IL 60439, USA}
\affiliation{Kavli Institute for Cosmological Physics, University of Chicago, Chicago, IL 60637, USA}

\begin{abstract}
We explore modifications to the loop-induced Higgs couplings $h\gamma\gamma$ and $h\gamma Z$ from light charginos in the Minimal Supersymmetric Standard Model.  When the lightest chargino mass is above the naive LEP bound of order $100$ GeV the effects are modest, with deviations in the decay branching ratios typically less than $15\%$ from the Standard Model predictions. However, if the charginos are lighter than $100$ GeV, more dramatic alterations to these couplings are possible as a consequence of the rise of the one loop form factor. For example, the diphoton signal strength can be enhanced by as much as $70\%$ compared to the Standard Model value. We scrutinize in detail the existing LEP, Tevatron, and LHC searches and present a scenario in which a very light chargino with a mass as light as half the Higgs mass is hidden at LEP and is allowed by all direct collider constraints and electroweak precision tests. The scenario has a sneutrino LSP with a macroscopic decay length of order $10 - 100$ cm. We outline potential search strategies to test this scenario at the LHC. 
\end{abstract}

\maketitle

The ATLAS and CMS experiments at the Large Hadron Collider have discovered a resonance at 125 GeV that appears in all respects to be the long-sought Higgs boson~\cite{ATLAS-Higgs, CMS-Higgs}. While this discovery represents an important milestone in our efforts to understand the mechanism of electroweak symmetry breaking, it also presents an exciting opportunity to probe physics beyond the Standard Model (SM) through precise studies of the properties of this new state. In this direction, there is now a dedicated experimental program to measure its couplings and quantum numbers. It is therefore of particular interest to understand the predicted range of Higgs couplings in motivated extensions of the Standard Model (SM), especially those that address the hierarchy problem such as the Minimal Supersymmetric Standard Model (MSSM). 

In this paper we revisit the modifications to the loop-induced couplings $h\gamma\gamma$ and $h\gamma Z$ from charginos in the MSSM. 
The Higgs-chargino-chargino coupling descends from the $SU(2)_L$ Higgs-Wino-Higgsino gauge interaction and thus has a magnitude of order $g/\sqrt{2}$, small in comparison to the top Yukawa coupling. Because of this fact, one generally expects that charginos lead to small deviations in the loop-induced couplings with respect to the SM predictions. This expectation was confirmed recently in Ref.~\cite{Casas:2013pta}. 

This conclusion, however, rests on the assumption that the chargino is heavier than the naive kinematic LEP bound of order 100 GeV. In this paper we will relax this assumption and entertain the possibility of charginos lighter than 100 GeV, and perhaps as light as $m_h/2$. In this regime, there is a strong enhancement in the one loop form factors governing the $h\rightarrow \gamma\gamma$, $h\rightarrow \gamma Z$ decays that can compensate the smallness of $g$. We will demonstrate that enhancements as large as 70$\%$ in the partial decay rates are then possible. While our focus is on the well-motivated case of charginos in the MSSM, the rise of the form factor is a generic feature for any new light charged particles coupled to the Higgs. 

As we will review in detail, the LEP experiments performed a suite of searches that are sensitive to chargino pair production, resulting in a fairly robust bound on the chargino mass that is close to the kinematic limit of order 100 GeV. However, as stressed recently in Ref.~\cite{Graham:2012th}, these searches have very weak coverage for neutral particles with a displaced decay length of order 10-100 cm due to strict impact parameter requirements for charged particles. 
Exploiting this fact, we present a scenario containing a very light chargino $\tilde \chi^+$, with mass $m_h/2 < m_{\tilde \chi^+} \lesssim 100$ GeV, that is allowed by all direct searches at LEP, Tevatron, and LHC. In our scenario, the LSP is a sneutrino with a macroscopic lifetime that decays within the detector through a small R-parity violating coupling. We also show that such light charginos can give an acceptable description of the precision electroweak data. Finally, we describe several potential experimental strategies to test this scenario at the LHC.

A great deal of theoretical effort has been focused on possible new physics contributions to the coupling of the Higgs to two photons. For example, besides the very light charginos that we discuss in this work, in the MSSM one can obtain a sizable enhancement in the $h\rightarrow \gamma\gamma$ rate with light, highly mixed staus~\cite{Carena:2011aa,Carena:2012gp,Carena:2012mw,Carena:2013iba}. These investigations have been motivated by data.  Early on in the search for the Higgs particle both ATLAS and CMS consistently reported measurements of the diphoton signal strength that were larger than one. While ATLAS still observes an enhancement in this channel, the latest CMS study no longer reports an enhancement. The most up-to-date measurements of the diphoton signal strength as of SUSY 2013 read~\cite{ATLAS-LP},\cite{CMS-LP}:
\begin{eqnarray}
\mu_{\gamma\gamma}  & = &  1.55^{+0.33}_{-0.28}, ~~~~~~~~ {\rm ATLAS}  ~~~\\  
\mu_{\gamma\gamma} &  =  & 0.78 \pm 0.27, ~~~~~  {\rm CMS}  
\end{eqnarray}
The quoted CMS result is based on a MVA analysis;  a cut based analysis is also performed and yields a higher central value: $\mu_{\gamma\gamma}   =   1.11 \pm 0.31$~\cite{CMS-LP}. The situation is therefore not clear experimentally, with the ATLAS and CMS disagreeing at the $2\sigma$ level. Thus, while there is certainly no statistically significant evidence for a large enhancement in this channel, there is still ample room for deviations to show up with the data accumulated during the upcoming 13 TeV run. If an unambiguous deviation in $h\rightarrow \gamma \gamma$ (or $h\rightarrow \gamma Z$) is eventually established, then it will be important to revisit with experiment the possibility of charginos and other hypothetical charged particles below $\sim 100$ GeV, since in this regime significant deviations are obtained even for weak couplings to the Higgs.

\section{Chargino contributions to \\ $h \rightarrow \gamma \gamma$ and $h \rightarrow \gamma Z $ }

We begin our study with an exploration of the possible range of corrections from very light charginos to the $h \rightarrow \gamma \gamma$ and $h \rightarrow \gamma Z $ rates. The one-loop chargino contributions to these rates have been computed in 
Refs.~\cite{Kalyniak:1985ct,Bates:1986zv,Gunion:1988mf,Weiler:1988xn,Djouadi:1996pb,Djouadi:1996yq}. 
These contributions have been incorporated into the {\sc CPSuperH} package for MSSM Higgs phenomenology, and formulae for the partial decay widths are collected in Refs.~\cite{Lee:2012wa,Lee:2007gn}. For the numerical results below, we have recalculated the partial decay rates and cross checked the results with {\sc CPSuperH}.

As alluded to above, one does not generally expect large contributions from chargino loops to these processes, as the $h^0\tilde \chi^+ \tilde \chi^-$ coupling is governed by the weak gauge coupling $g$. We can see this explicitly by employing the low-energy theorem~\cite{Ellis:1975ap,Shifman:1979eb} to compute the effective $h\gamma\gamma$ coupling, which in the decoupling limit is given by 
\begin{equation} 
{\cal L} \supset \frac{\alpha \, b_{\tilde \chi^+} }{16 \sqrt{2} \pi v} \!
 \left[ \!  \left(   \frac{\partial}{\partial \log v_u} \! + \!  \frac{\partial}{\partial \log v_d}     \right) 
 \! \log {\bf X}{\bf X}^\dag \!   \right] h^0 F_{\mu\nu}F^{\mu\nu}, ~
 \label{LET}
\end{equation}
where $v_u, v_d$ are the vacuum expectation values for the up- and down-type Higgs, $v \equiv \sqrt{v_u^2 + v_d^2} = 174$ GeV, $b_{\tilde \chi^+} = 4/3$ is QED beta function coefficient for the chargino, and ${\bf X}$ is the chargino mass matrix, 
\begin{equation}
{\bf X} = \left(  
\begin{array}{cc}
M_2   &   g v_u \\
g v_d & \mu
\end{array}
   \right), 
\end{equation}
with $M_2$ the wino soft mass and $\mu$ the supersymmetric Higgs mass parameter. 
We can straightforwardly take the derivatives in Eq.~(\ref{LET}) 
and compute the correction to the partial decay width: 
\begin{equation}
\frac{ \Gamma(h\rightarrow \gamma\gamma)}{ \Gamma(h\rightarrow \gamma\gamma)_{\rm SM}} \simeq
\bigg\vert   1 + \frac{b_{\tilde \chi^+}}{A_{\gamma}^{\rm SM}  } \frac{g^2 v^2 \sin {2 \beta} }{M_2 \mu - \tfrac{1}{2} g^2 v^2 \sin{2\beta}}    \bigg\vert^2,~~
\label{rgamma-LET}
\end{equation} 
where $\tan\beta \equiv v_u/v_d$ and $A_{\gamma}^{\rm SM} \approx 6.5$ comes mainly from the $W$ and top loop. We observe that the chargino contributions is maximized for $\tan \beta \rightarrow 1$. However, even in this case, if the lightest chargino is above $\sim 100$ GeV, only a moderate contribution to the $h\rightarrow \gamma \gamma$ partial width is achieved. For example, with $\tan\beta = 1$, $M_2 = \mu = 185$ GeV, we obtain the chargino mass eigenvalues $(m_{\tilde \chi^+_1},m_{\tilde \chi^+_2}) =(104, 266)$ GeV and an enhancement in the $h^0 \rightarrow \gamma\gamma$ rate of $\sim 20\%$ from Eq.~(\ref{rgamma-LET}). In this case, the lightest chargino is just above the LEP limit  of $103.5$ GeV assuming the decay $\tilde \chi_1^+ \rightarrow W^+ \tilde \chi^0$ and R-parity conservation resulting in a stable neutralino $\tilde \chi^0$. This is the maximum enhancement possible for charginos heavier than the LEP bound, and the effect decreases dramatically as the lightest chargino mass is raised and as $\tan\beta$ is increased. 
\begin{figure}
\centerline{
\includegraphics[width= 0.9\columnwidth]{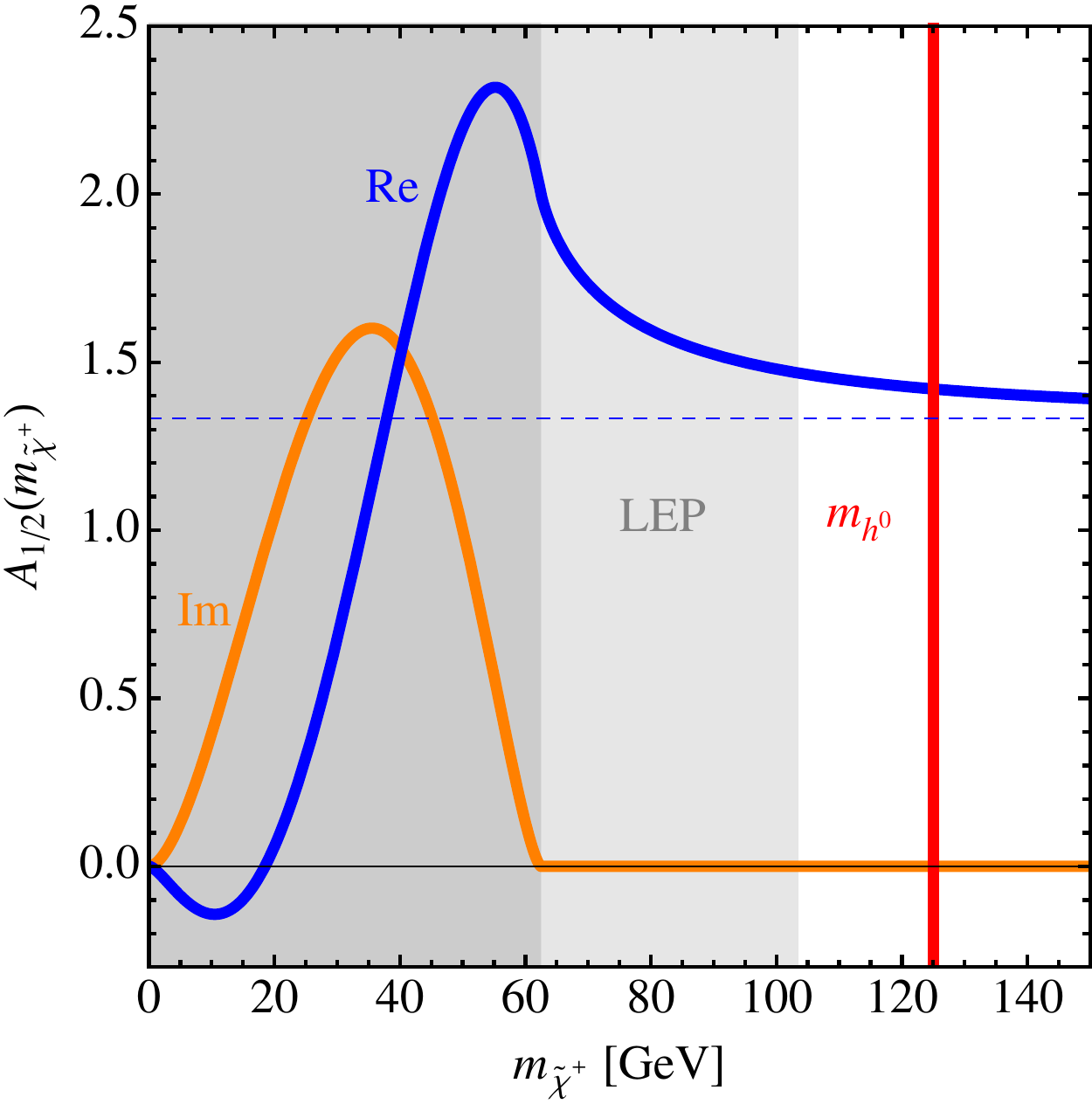}
}
\caption{Form factor $A_{1/2}(m_{\tilde \chi^+})$ for $h \rightarrow \gamma\gamma$ mediated by a chargino loop (our definition of $A_{1/2}$ matches that of Ref.~\cite{Djouadi:2005gi}). We display the real (blue) and imaginary (orange) parts of the form factor. The red line indicates the Higgs mass. The dark shaded region $m_{\tilde \chi^+} < m_h/2$ is not allowed by the Higgs signal strength dataset, since for such light charginos the Higgs would dominantly  decay to chargino pairs. Direct searches at LEP place model dependent constraints in the light shaded region $m_{\tilde \chi^+} \lesssim 100$ GeV. As the mass of the chargino is lowered, the form factor increases from $A_{1/2} = 4/3$ for very heavy charginos to $A_{1/2} = 2$ for $m_{\tilde \chi^+} = m_h/2$. } 
\label{fig:form-factor}
\end{figure}

We now relax the assumption that the chargino is heavier than 100 GeV and consider lighter charginos. We restrict to values of the lightest chargino mass eigenstate greater than $m_h/2$ since otherwise the Higgs boson will dominantly decay to a pair of charginos, which is not allowed by the Higgs signal strength dataset. Later we will present a concrete scenario in which such a light chargino evades all direct searches at LEP, but the chargino contributions to the loop-induced Higgs couplings are independent of these considerations. The point we wish to emphasize here is very simple: as the mass of the lightest chargino $m_{\tilde \chi_1^+}$ decreases towards $m_h/2$, the one loop form factor $A_{1/2}(m_{\tilde \chi^+_1})$ exhibits a rise,  as displayed in Fig.~\ref{fig:form-factor}. The form factor increases from its asymptotic value of $A_{1/2} = 4/3$ for heavy charginos to $A_{1/2} = 2$ for $m_{\tilde \chi_1^+} = m_{h}/2$.  This basic observation, which of course applies not only for charginos but for all light charged particles that couple to the Higgs, allows for a large correction to the loop-induced couplings. 

\begin{figure}[t!]
\centerline{
\includegraphics[width= 0.9\columnwidth]{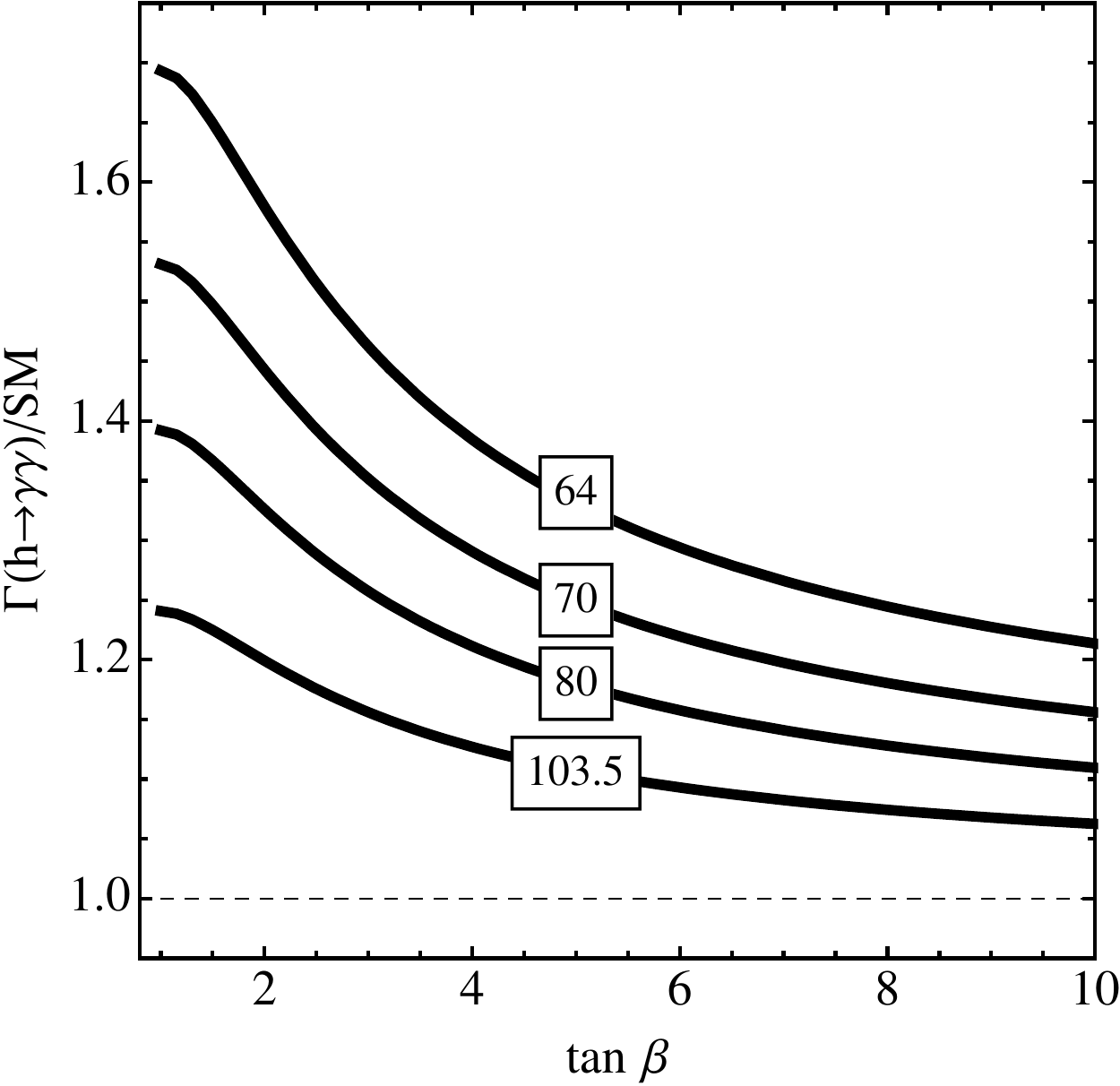}
}
\caption{Ratio of the partial decay width for $h\rightarrow \gamma\gamma$  to the SM prediction including the contribution of charginos. We have fixed $M_2 = \mu$ and work in the decoupling limit. The prediction is shown for lightest chargino masses $m_{\tilde \chi^+_1} = 64, 70, 80, 103.5$ GeV.} 
\label{fig:hGaga}
\end{figure}

\begin{figure}[t!]
\centerline{
\includegraphics[width= 0.9\columnwidth]{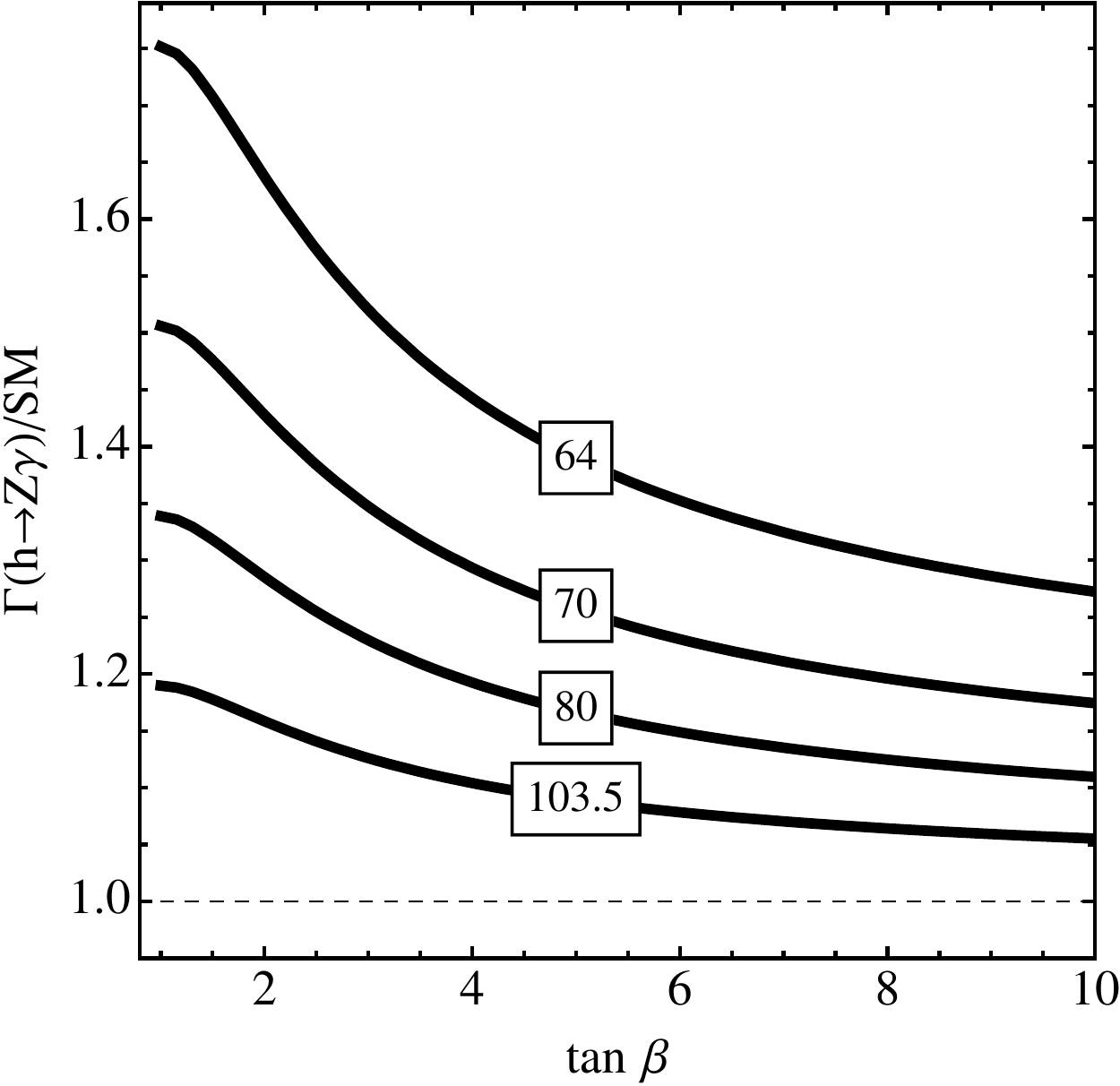}
}
\caption{ Ratio of the partial decay width for $h\rightarrow \gamma Z$  to the SM prediction including the contribution of charginos. We have fixed $M_2 = \mu$ and work in the decoupling limit. The prediction is shown for lightest chargino masses $m_{\tilde \chi^+_1} = 64, 70, 80, 103.5$ GeV. } 
\label{fig:hGaZ}
\end{figure}

We present in Figs.~(\ref{fig:hGaga}) and (\ref{fig:hGaZ}) the ratios of the partial decay widths to their SM values, including the one-loop chargino contributions, for $h\rightarrow \gamma\gamma$ and $h\rightarrow Z \gamma$, respectively. The results are displayed as a function of $\tan \beta$ for several values of the lightest chargino mass eigenstate between $m_h/2$ and the LEP limit. In this plot we have fixed $M_2 = \mu$, which maximizes the chargino contribution. We observe that it is possible to obtain enhancements as large as $70\%$ (75$\%$) for the $h\rightarrow \gamma\gamma$ ($h\rightarrow \gamma Z$) rates in the extreme case of $m_{\tilde \chi^+_1} = 64$ GeV and $\tan \beta = 1$. However, even for somewhat larger chargino masses and $\tan \beta$ values, it is possible to achieve a sizable enhancement. For instance, if  $m_{\tilde \chi^+_1} = 70$ GeV and $\tan \beta = 4$, one obtains enhancements of order $30\%$.

We note that the LHC can eventually measure the $h\gamma\gamma$ coupling with a precision of $5-10\%$ with a high luminosity $3000\, {\rm fb}^{-1}$ run, while a future linear collider can achieve a precision of a few percent~\cite{Peskin:2012we},\cite{Klute:2013cx}. In contrast to the $h \rightarrow \gamma \gamma$ channel, very few detailed studies exist for the $h \rightarrow \gamma Z$ channel. The analyses of Refs.~\cite{Gainer:2011aa},\cite{Campbell:2013hz} indicate that the SM prediction for this channel can be probed with $O(100\,{\rm  fb}^{-1})$ at the LHC, but as far as we are aware no projections for future colliders exist.

\section{Collider bounds on charginos}

While we have speculated in the previous section on the existence of a light chargino with mass $m_h/2 < m_{\tilde \chi^+} \lesssim 100$ GeV, the conventional wisdom states that charginos are ruled out up to the kinematic reach at LEP2, $m_{\tilde \chi^+} \gtrsim 100$ GeV. We now survey the existing limits from LEP on such light charginos, the assumptions going into such limits, and potential loopholes in these analyses. We will break the discussion into scenarios with and without R-parity conservation. In the next section we will present a concrete scenario with a chargino as light as $m_h/2$ which evades all direct collider constraints.

\paragraph{\bf R-parity conservation} 
We first consider scenarios with a conserved R-parity and stable LSP.
\begin{itemize}
\item {\it Chargino LSP}
\footnote{In the MSSM the neutralino is generically lighter than the chargino, although an exception to this rule occurs when 
${\rm sign}(M_1) \neq {\rm sign}(M_2) = {\rm sign}(\mu)$. 
See Ref.~\cite{Kribs:2008hq} for a detailed analysis.}:  A stable (or long-lived) chargino LSP is ruled out up to the kinematic limit at LEP by searches for charged massive stable particles, {\it e.g.}, 
Refs.~\cite{Abdallah:2002rd,Abbiendi:2003yd}. The actual bound is much higher in chargino mass by now due to similar searches at the Tevatron and LHC. Beyond direct collider searches, strong bounds on cosmologically stable charginos exist from from searches for heavy water isotopes, see, {\it e.g.}, Ref.~\cite{Yamagata:1993jq}.
\item {\it Neutralino LSP, Chargino NLSP}: This is the canonical scenario for a light chargino. There are two cases to consider: 

1) In the case of sizable chargino-neutralino mass splitting $\Delta M$, the limit on the chargino mass is 103.5 GeV~\cite{neutralino-chargino-RPC}. This limit assumes that the sneutrinos are heavy, $m_{\tilde \nu} > 300$ GeV. If a light electron-type sneutrino is present in the spectrum, $m_{\tilde \nu_e} \lesssim 100$ GeV, the chargino pair production cross section can be as small as $\sim 0.5$ pb (for $M_2 = \mu$) depending on the parameters. The upper limit on the chargino pair production cross section~\cite{neutralino-chargino-RPC} is larger than this in some regions of parameters space, particularly for chargino masses  $ m_{\tilde \chi_1^+} \gtrsim 65$ GeV and light neutralinos, $ m_{\tilde \chi_1^0} \lesssim 20$ GeV. A detailed investigation of this possible opening is beyond the scope of this paper, but we note that there are several important additional considerations besides the LEP cross section limit: 1) constraints from Higgs decays to neutralinos, 2) implications of a light sneutrino (we will discuss this issue below), 3) collider searches from Tevatron and LHC (the chargino/neutralino production cross section at hadron colliders does not depend on the properties of the sneutrino).
 
2) In the case of a small chargino-neutralino mass splitting $\Delta M$, dedicated searches were carried out to cover this region~\cite{neutralino-chargino-RPC-small-DM}. The limits are quite strong in general, but there is a region from $0.1\,{\rm GeV} \lesssim \Delta M \lesssim 3\,{\rm GeV}$  where the limit is slightly weaker than the naive kinematic limit, $m_{\tilde \chi^+_1} \gtrsim 92$ GeV. We note that in this case, one can obtain a maximum enhancement of about $30\%$ in the $h\rightarrow \gamma\gamma$ decay rate from chargino loops. 
\item {\it Sneutrino LSP - Chargino NLSP}:  In this case, the chargino will decay to a lepton and a stable sneutrino, leading to a slepton-like signature of acoplanar leptons plus missing momentum. The combined LEP2 upper limits on the slepton production cross section are very strong in this case, below 0.2 pb for the case of stau, and much lower for smuon and selectron~\cite{slepton}. There is a possible loophole in the region of small mass splitting between the slepton (or chargino) and LSP, as in this regime the leptons are very soft and the missing momentum is small as the LSPs travel back-to-back. 
For example, the ALEPH slepton search~\cite{Heister:2001nk} requires each lepton to have $p_T > 0.5\%\sqrt{s}$ and missing transverse momentum $p_{T \rm miss} > 1\% \sqrt{s}$, criteria which will generically not be met for splittings smaller than a GeV. We note that, unlike a neutralino and chargino, there is no particular reason that a chargino and a sneutrino should be so close in mass, but it is nevertheless an interesting possibility.
\end{itemize}

\paragraph{\bf R-parity violation}
We now consider scenarios with R-parity violation, in which the LSP is unstable. The signatures of chargino pair production are rich and varied in this case, with multiple leptons, jets and missing energy (from neutrinos)  as a possible final states. If the chargino is the LSP, it will decay to three SM fermions:  2 charged leptons and a neutrino through $LLE$, 2 jets and a lepton or neutrino via $LQD$, and three jets via $UDD$. If the chargino is the NLSP, or there are additional light superpartners in the spectrum, the final state multiplicity is even higher. The LEP experiments performed a broad array of searches to cover this diversity of signatures.

The ALEPH~\cite{Heister:2002jc}, DELPHI~\cite{Abdallah:2003xc}, and OPAL~\cite{Abbiendi:2003rn} analyses present {\it inclusive} limits from multiple search channels interpreted within the framework of the MSSM in the $\mu-M_2$ parameter space; the chargino mass limit is $m_{\tilde \chi^+_1} \gtrsim 100$ GeV in all cases. In contrast, the L3 analysis~\cite{Achard:2001ek} presents limits on specific decay modes for assumed spectra and RPV couplings, which are more easily interpreted. Again, charginos  are ruled out up to $\sim 100$ GeV in all cases, though not all RPV couplings are considered.  

For the ALEPH, DELPHI, and OPAL analyses, since explicit limits on specific decay modes of the chargino are not presented, it is  difficult to make a concrete statement on the mass limit of the chargino for a specific spectrum and RPV coupling  without further detailed analysis and simulation to recast these results. Nonetheless, in all of the RPV searches referenced above the observed number of events and expected backgrounds are reported for each selection, and no statistically significant excess was observed. Given the large chargino pair production cross section, one expects that the upper limit on the chargino mass will be close to the kinematic reach for any conceivable decay path. 

\section{Chargino hiding at LEP} 

As we have seen, a variety of searches at LEP place a fairly robust bound on charginos close to the naive kinematic limit, $m_{\tilde \chi^+_1}  \gtrsim 100~\GeV$. 
An explicit assumption in the RPV searches is that the LSP decays promptly. 
However, as emphasized recently in Ref.~\cite{Graham:2012th}, these bounds are often severely weakened or even nullified if the LSP has a displaced decay within the detector. 
In this section we describe a viable scenario in which the lightest chargino has a mass $m_h/2 < m_{\tilde \chi^+_1}  \lesssim 100~\GeV$. In our scenario, the LSP is a  sneutrino that has a macroscopic decay length of order $10 - 100~{\rm cm}$.
We will demonstrate that this scenario is compatible with all existing direct searches by experiments at LEP, Tevatron, and LHC.

\paragraph{\bf Long-lived Neutral LSP at LEP} 
We begin by describing how long-lived neutral states decaying to jets and/or charged leptons can evade the suite of searches performed at LEP. We may first ask how such objects are classified during event reconstruction. For concreteness, let us focus on the ALEPH experiment~\cite{Buskulic:1994wz}, which uses an energy flow algorithm. The process begins with the identification of ``good'' charged tracks. Any charged track which originates from within a cylinder of length 20 cm and coaxial radius 2 cm from the beam line and centered about the interaction point will be considered as a ``good'' track. 
Crucially, for a long-lived neutral particle with a macroscopic lifetime, $c \tau \gtrsim 10$ cm, the tracks associated with the charged decay products will be ignored as they will generally not point back to the collision point.  The LEP RPV searches typically require a minimum number of ``good'' tracks, a significant fraction of which are assumed to originate from the prompt decay of the LSP. This selection criterion will generally not be met if the LSP decays far away from the primary interaction region. Similar conclusions hold for the DELPHI, OPAL, and L3 experiments~\cite{Abdallah:2003xc,Abbiendi:2003rn,Achard:2001ek}. 

What then becomes of the particles produced in the displaced decay of the long-lived neutral particle? If the decay is displaced but still within the detector, the final state jets and/or leptons will deposit energy in the hadronic and/or  electromagnetic calorimeters, and as such the long-lived neutral particle will not leave a signature of missing transverse momentum.  Instead,  these particles will be classified as neutral hadrons by the energy flow algorithm~\cite{Buskulic:1994wz}. 

For lifetimes greater than several meters, the LSP can escape the detector altogether, leading to events with missing transverse momentum. Thus, in this regime standard SUSY searches will place severe constraints on sparticle masses. However, as we will show, there is an important gap for lifetimes of ${\cal O}(10 - 100\,{\rm cm})$ for which a displaced neutral LSP can evade the searches at LEP.

\paragraph{\bf Very Light Chargino}

We now describe a viable scenario for a very light chargino with mass $m_h/2 < m_{\tilde \chi^+_1}  \lesssim 100~\GeV$. For concreteness, we take the electron-type sneutrino to be the LSP and assume it decays via the $\lambda_{121}$ $LLE$ coupling to an electron and anti-muon,
\begin{equation}
\tilde \nu_e \rightarrow e^- + \mu^+.
\label{eq:sneutrino-decay}
\end{equation}
The partial decay width is given by 
$\Gamma_{\tilde \nu_e \rightarrow e^- \mu^+} \simeq \lambda_{121}^2 m_{\tilde \nu_e}/16\pi$, implying a sneutrino decay length
\begin{equation}
c\tau_{\tilde \nu_e} \approx 1\,{\rm  cm} \times \left(\frac{10^{-7}}{\lambda_{121}} \right)^2 \left( \frac{70\,~\rm GeV}{m_{\tilde \nu_e}}\right).
\end{equation}
Such a small RPV coupling of order $10^{-7}$ is allowed by all experimental constraints~\cite{Barbier:2004ez}. Furthermore, constraints from lepton flavor violation are model dependent and can be evaded with appropriate choices of the slepton and sneutrino soft masses and trilinear couplings.  
In addition to the sneutrino and chargino, there will necessarily be a light selectron and neutralino in the spectrum, and we will also consider the constraints on these states as well. Although our quantitative results apply to the specific case of an electron sneutrino LSP decaying according to (\ref{eq:sneutrino-decay}), there are other viable possibilities for the flavor and decay mode of the sneutrino. We will comment on these and other possibilities for the LSP and its decays below. 

We now carefully consider the existing constraints on this scenario from direct collider searches and derive exclusion regions as a function of the sneutrino lifetime.  
To obtain our numerical results we have used  {\sc{MadGraph5}}~\cite{madgraph} for event generation, implementing the RPV-MSSM model with {\sc{FeynRules}}~\cite{Christensen:2008py},  in conjunction with a private code to to compute the efficiency of track selection accounting for the finite decay length of the sneutrino.

\paragraph{ Sneutrino LSP}

We first consider the constraints on direct pair production of the sneutrino LSP at LEP. With the sneutrino decay mode in Eq.~(\ref{eq:sneutrino-decay}),  pair production will lead to a four lepton final state. The ALEPH search~\cite{Heister:2002jc} places a lower bound on the $\tilde \nu_e$ mass close to the kinematic limit, $m_{\tilde \nu_e} \gtrsim 100$ GeV, in the case when the sneutrino decays promptly. However, the analysis requires 4 good tracks~\cite{Barate:1997ra}, and this requirement will generally not be met for sneutrinos with a sufficiently displaced decay. 

\begin{figure}
\centerline{
\includegraphics[width=0.9\columnwidth]{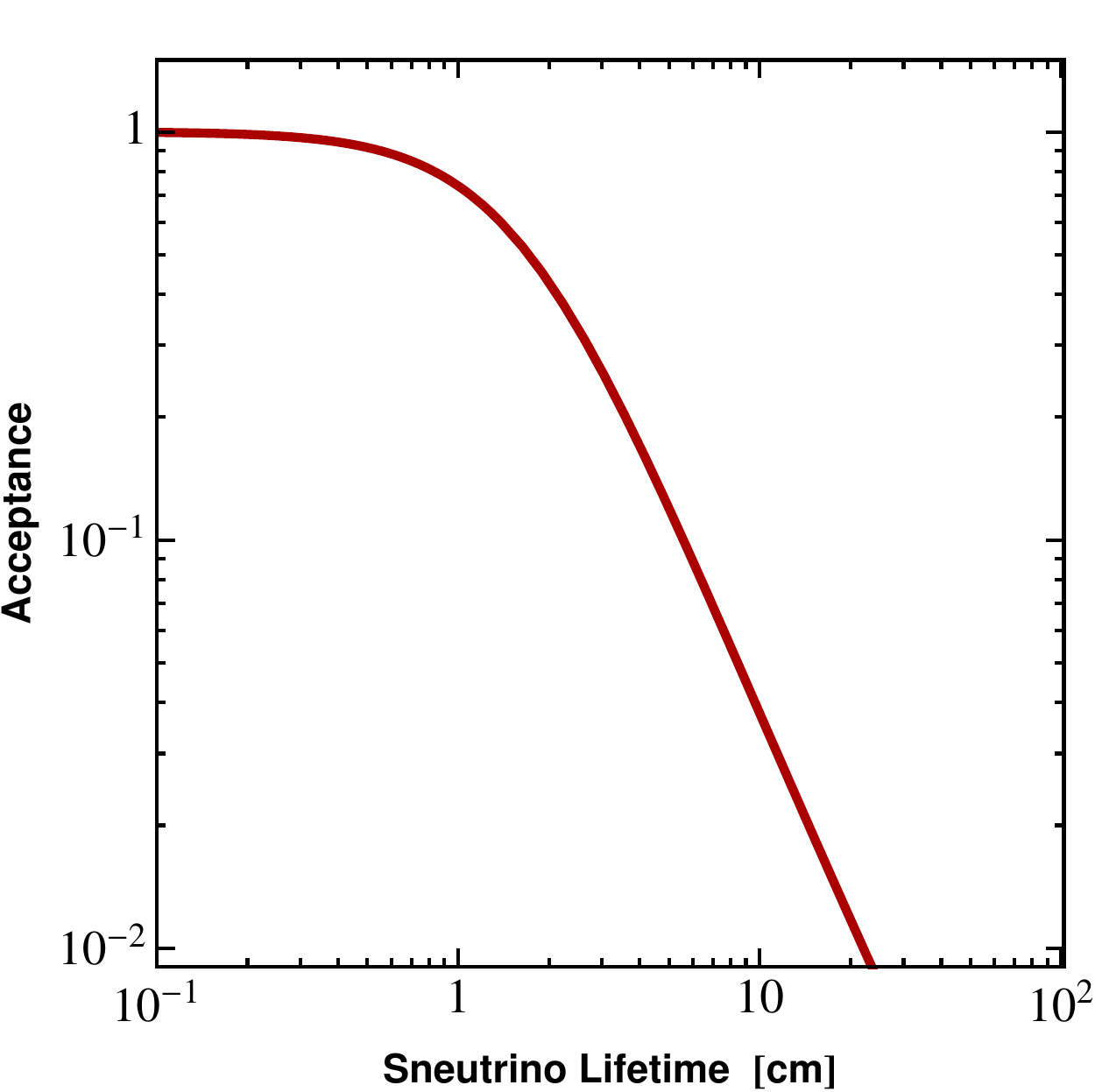}
}
\caption{
Efficiency to select events with four good tracks in sneutrino pair production followed by $\tilde \nu_e \rightarrow e^- \mu^+$ for $\sqrt{s} = 208$ GeV as a function of the sneutrino lifetime $(c \tau_{\tilde \nu_e})$. Here we have fixed $m_{\tilde \nu_e} = 70$ GeV. 
 }
\label{fig:sneutrino}
\end{figure}

In Fig.~\ref{fig:sneutrino} we display the fraction of signal events accepted as a function of $c \tau_{\tilde \nu_e}$ after demanding that each lepton points back to a cylinder of length 20 cm and coaxial radius 2 cm centered about the collision. We observe that the acceptance decreases dramatically for sneutrino lifetimes larger than the cylinder radius, as in this regime the leptons do not typically point back to the cylinder. 

\begin{figure}
\centerline{
\includegraphics[width= 0.9\columnwidth]{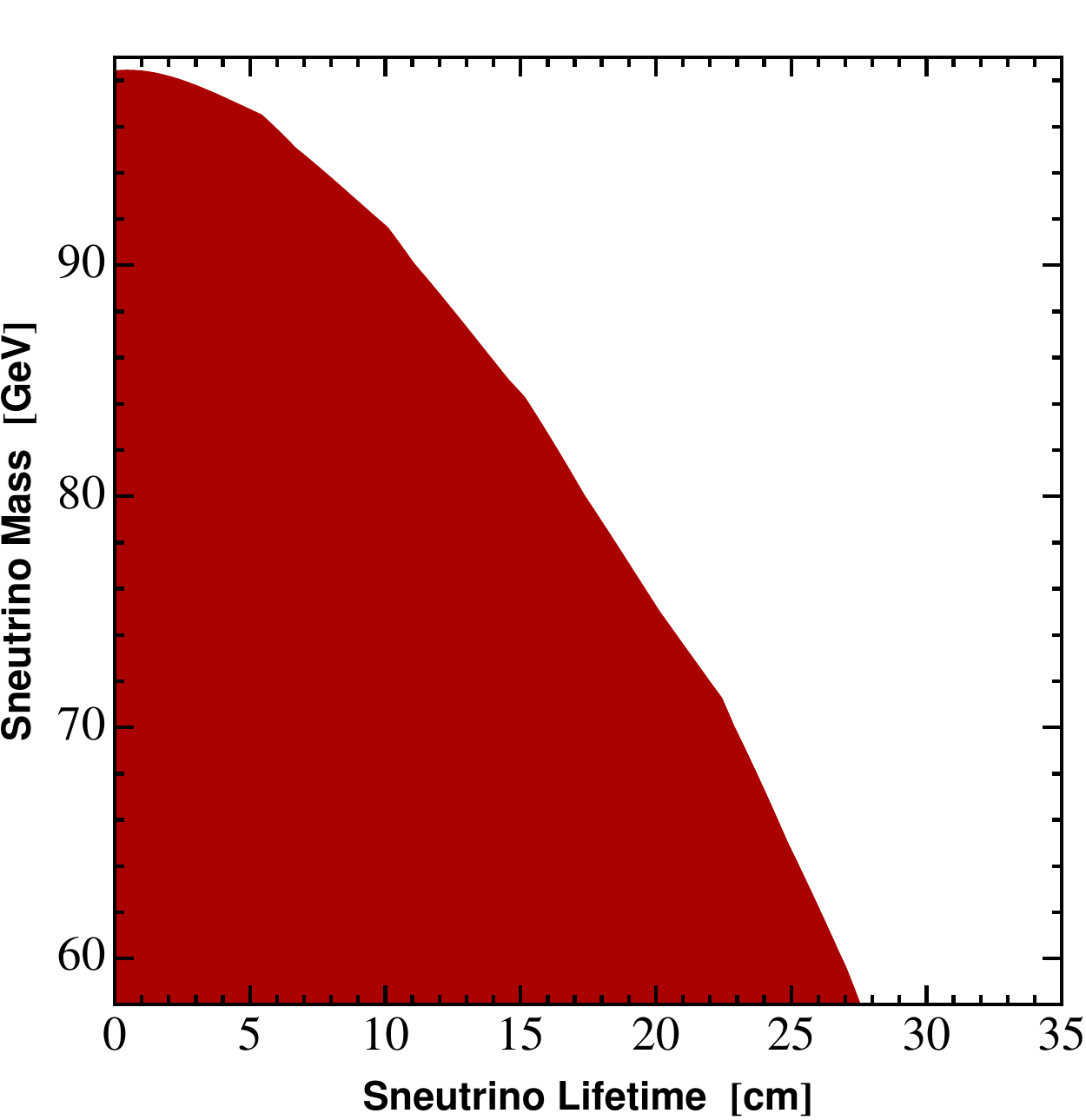}
}
\caption{Constraints from ALEPH RPV four lepton searches~\cite{Heister:2002jc} on an electron sneutrino LSP decaying via $\tilde \nu_e \rightarrow e^- \mu^+$  in the sneutrino lifetime ($c\tau_{\tilde \nu_e}$) - sneutrino mass plane. Here we have fixed $m_{\tilde \chi_1^+} = m_{\tilde \nu_e} + 10$ GeV,  $M_1 = \mu$, and $\tan\beta = 1.5$
} 
\label{fig:sneutrino-bound}
\end{figure}

In Fig.~\ref{fig:sneutrino-bound} we present the bounds from this search in the sneutrino lifetime ($c\tau_{\tilde \nu_e}$) - sneutrino mass plane. To derive these bounds, we equate the product of the sneutrino pair production cross section and the efficiency to select four good tracks 
to the RPV four lepton cross section limit in Fig.~6a of Ref.~\cite{Heister:2002jc}. 
The sneutrino pair production cross section can be found in Ref.~\cite{Dreiner:2008tw}. It is important to note that for electron sneutrino pair production in $e^+ e^-$ reactions, there is an additional diagram involving the $t$-channel exchange of the chargino. We have made the assumption $m_{\tilde \chi_1^+} = m_{\tilde \nu_e} + 10$ GeV,  
$M_1 = \mu$, and $\tan\beta = 1.5$ in Fig.~\ref{fig:sneutrino-bound}. 
In fact, the chargino diagram dominates in this case, leading to a cross section as large as $3$ pb for $m_{\tilde \nu_e} = 65$ GeV; the bound on a muon or tau flavored sneutrino would be much weaker. Nevertheless, we conclude from Fig.~\ref{fig:sneutrino-bound} 
that electron sneutrino LSPs  decaying to $e\mu$ pairs with decay lengths of 25 cm or longer are not constrained by this search.  We note that Ref.~\cite{Graham:2012th} also derived bounds on a long-lived sneutrino LSP for the case in which it decays to a $b\bar b$ pair. 

What is the lower limit on the sneutrino mass? Sneutrinos lighter than $m_Z/2$ will be constrained by the total width of the $Z$ boson. Furthermore, the Higgs can decay to sneutrino pairs if 
$m_{\tilde \nu_e} < m_h/2$. The coupling of the Higgs to sneutrinos is 
$\lambda_{h^0 \tilde \nu \tilde \nu^*} = m_Z^2 \cos{2\beta}/\sqrt{2} v$, and thus $h\rightarrow \tilde \nu_e \tilde \nu_e^*$ would be the dominant decay mode for all values of $\tan \beta $ except those very close to unity.
Up to the additional phase space suppression for $m_{\tilde \nu_e} \lesssim m_h/2$,
we find that for $\tan \beta  \lesssim 1.2$, the Higgs branching fraction to sneutrino pairs is 
${\rm Br}(h\rightarrow \tilde \nu_e \tilde \nu_e^*) \lesssim 0.2$ which is allowed by current LHC data on the Higgs signal strength measurements. However, it is important to emphasize that an additional decay mode of the Higgs to sneutrino pairs will also dilute any enhancement in $h\rightarrow \gamma \gamma, \gamma Z$  due to the light chargino.  We will therefore restrict to sneutrino masses greater than $m_h/2$. 

\paragraph{Light Chargino}

Given that the electron sneutrino LSP can be lighter than 100 GeV provided its decay is displaced, $c \tau_{\tilde \nu_e } \gtrsim 25$ cm, we now investigate whether a very light chargino with mass $m_h/2  < m_{\tilde \chi^+} \lesssim 100$ GeV is also allowed in this scenario. The chargino will decay promptly via 
\begin{equation}
\tilde \chi_1^+ \rightarrow   e^+ +  \tilde \nu_e, 
\label{charginodecay}
\end{equation}
followed by the displaced sneutrino decay (\ref{eq:sneutrino-decay}) to a $e^- \mu^+$ pair. Chargino pair production will thus lead to a six lepton final state, with four of the leptons being displaced. There are two classes of LEP searches that are sensitive to chargino pair production in this case. 
The first class of searches are for multi-lepton final states, as in RPV searches~\cite{Heister:2002jc,Abdallah:2003xc,Abbiendi:2003rn,Achard:2001ek}. These searches will be relevant for prompt sneutrino decays. However, as in the case of direct sneutrino pair production discussed above, the multi-lepton RPV searches explicitly require a minimum number of good tracks and their reach will be significantly diminished as the sneutrino lifetime increases. 
The second class involves searches for acoplanar leptons plus large missing transverse momentum~\cite{Heister:2001nk,Abdallah:2003xe,Achard:2003ge,Abbiendi:2003ji}, which is the canonical signature of sleptons in R-parity conserving scenarios, and is relevant for very long sneutrino lifetimes such that the sneutrino decays outside the detector.

\begin{figure}
\centerline{
\includegraphics[width= 0.93\columnwidth]{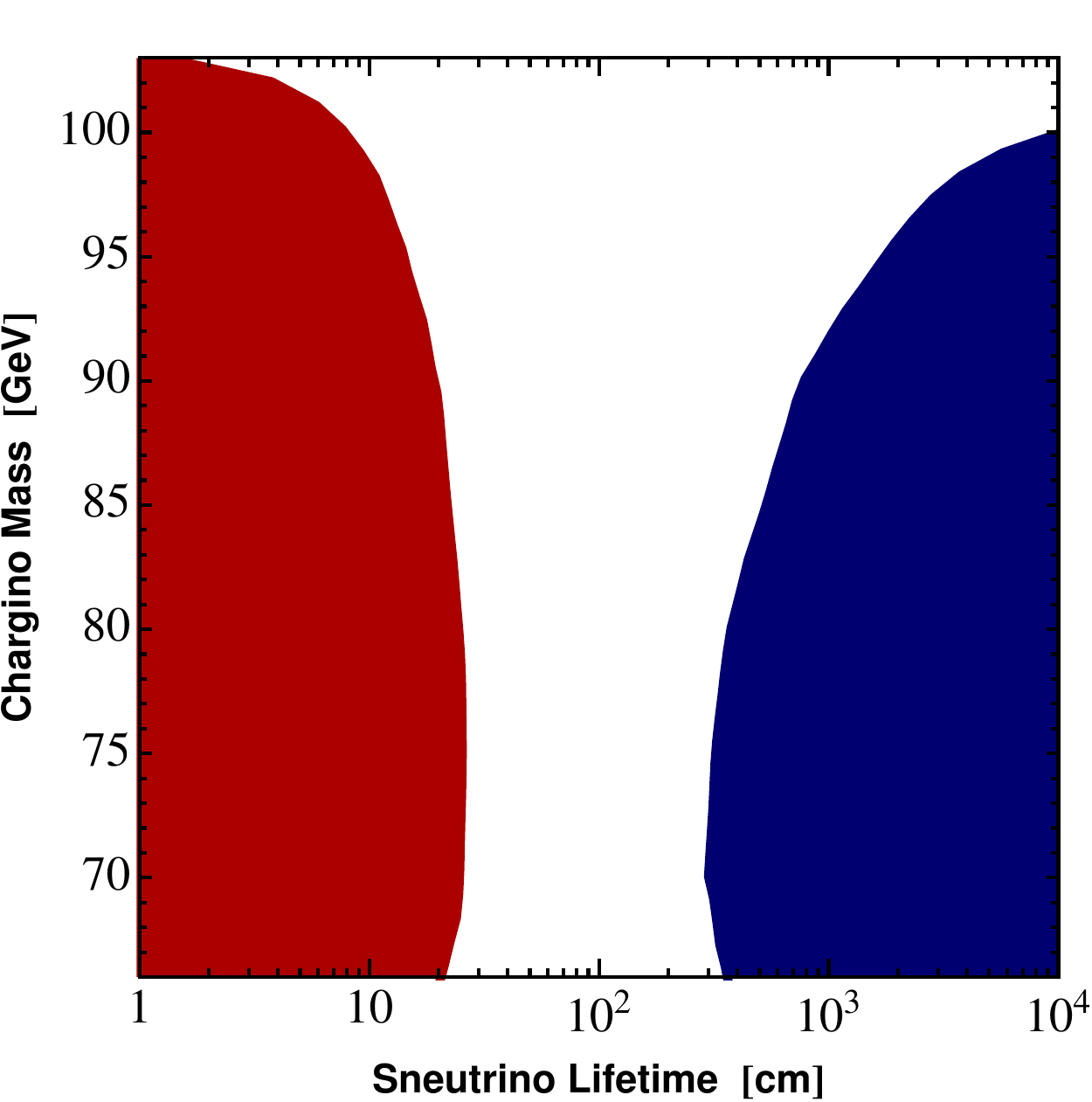}
}
\caption{Constraints in the chargino mass - sneutrino lifetime plane on chargino pair production 
with the subsequent  decay chain $\tilde \chi_1^+ \rightarrow   e^+ +  \tilde \nu_e$,  $\tilde \nu_e \rightarrow e^- + \mu^+$. The red shaded region is ruled out by the OPAL six lepton RPV search~\cite{Abbiendi:2003rn}, while the blue shaded region is ruled out by the ALEPH slepton search (acoplanar leptons plus missing transverse momentum)~\cite{Heister:2001nk}. We have fixed $m_{\tilde \nu_e} = 65$ GeV, $M_2 = \mu$, and $\tan\beta = 1.5$.} 
\label{fig:chargino-bound}
\end{figure}

We present in Fig.~\ref{fig:chargino-bound} the bounds on chargino pair production as a function of the chargino mass and sneutrino LSP lifetime. First, the red shaded region is constrained by the OPAL RPV six lepton search~\cite{Abbiendi:2003rn}. 
To derive these bounds, we equate the product of the sneutrino pair production cross section (see, {\it e.g.}, Ref.~\cite{Dreiner:2008tw}) and the efficiency to select at least 5 good tracks obtained through our simulation
to the six lepton cross section limit in Fig.~7 of Ref.~\cite{Abbiendi:2003rn} \footnote{While Fig.~7 of Ref.~\cite{Abbiendi:2003rn} presents a limit on slepton pair production resulting in six lepton events with large missing energy, we note that the search also included selections sensitive to six leptons without missing energy, and thus applies to our scenario.}. A ``good'' track for the OPAL search is one that points back to a cylinder with a 1 cm coaxial radius and 40 cm length around the primary collision~\cite{Abbiendi:2003rn}.  The presence of  an electron sneutrino LSP leads to an additional $t$-channel sneutrino exchange diagram in chargino pair production, which interferes destructively with the $s$-channel $\gamma,Z$ exchange amplitudes. For the limit in Fig.~\ref{fig:chargino-bound}  we have fixed $m_{\tilde \nu_e} = 65$ GeV, $M_2 = \mu$, and $\tan\beta = 1.5$. 

The blue shaded region in Fig.~\ref{fig:chargino-bound} is constrained by the ALEPH slepton search for acoplanar leptons plus missing transverse momentum~\cite{Heister:2001nk}. To place the bound we calculate as a function of $c \tau_{\tilde \nu_e}$ the efficiency for both sneutrinos emitted in the decays of the charginos to escape the detector before decaying. The chargino cross section scaled by this efficiency is then compared to the observed cross section limit on selectron pair production (Fig.~7a of Ref.~\cite{Heister:2001nk}). We note that the search~\cite{Heister:2001nk} employs a neutral veto to eliminate SM dilepton events with hard initial or final state radiation.
The invariant mass between any good charged track and neutral energy flow particle is required to be very small. If one of the sneutrinos has a displaced decay within the detector it will be classified as a neutral hadron, and the event will be cut as a result of the neutral veto. In our simulation, we therefore select events in which both sneutrinos decay outside the detector 

We note that for larger lifetimes there can be an additional source of missing transverse momentum due to the mis-measured momentum of the sneutrino. As the displaced sneutrinos in our scenario will be classified as neutral hadrons, it is likely their momentum is determined by tracing hits in the calorimeter back to the event vertex. This will result in a measured momentum that is different than their true momentum. As such, the total momentum of visible objects in the event will not balance. We have simulated this effect and find that for lifetimes of order 100 cm or more the missing momentum can be $\sim 5-10$ GeV, large  enough to pass the missing $p_T$ selection of the slepton search in Ref.~\cite{Heister:2001nk}. However, as discussed above, the neutral veto employed in this analysis requires both sneutrinos to escape the detector. Thus, the bound in Fig.~\ref{fig:chargino-bound} is not affected by this additional source of missing transverse momentum. See Ref.~\cite{Fan:2012jf} for  further discussion of the effects of mis-measured momentum. 

We conclude from Figs.~\ref{fig:sneutrino-bound} and \ref{fig:chargino-bound} that there is a large range of sneutrino LSP lifetime between $25\,{\rm cm} \lesssim c \tau_{\tilde \nu_e} \lesssim 300\,{\rm cm}$ for which light charginos with mass $m_{\tilde \chi^+_1} \gtrsim m_h/2$  are allowed by direct searches at LEP. 

\paragraph{Light neutralino and selectron}
Thus far we have discussed only the sneutrino LSP and the very light chargino below 100 GeV . However, in addition to these states, there will also necessarily be a light neutralino and light selectron in the spectrum. We now discuss the possible constraints on these light states from LEP searches. 

We first consider the lightest neutralino. To see that there is a light neutralino, consider the limit
 $M_2 = \mu$, $\tan \beta = 1$ relevant for large modifications to the $h\gamma\gamma$ and $h\gamma Z$ couplings. For $|M_1| \gg M_2$ we obtain the following approximate expressions for the mass eigenvalues of the chargino-neutralino system:
\begin{eqnarray}
M_{\tilde \chi^+}  &  \approx & \left(  M_2 -\frac{g v}{\sqrt{2} } , M_2 +\frac{g v}{\sqrt{2}} \right) \nonumber \\
 M_{\tilde \chi^0}  & \approx  &  \left(  M_2 -\frac{g v}{\sqrt{2}}, M_2, M_2 +\frac{g v}{\sqrt{2}} , |M_1|
 \right) 
\end{eqnarray}
One observes that the lightest chargino and neutralino are degenerate in this limit. Moving away from this limit, one still always finds a light neutralino eigenvalue accompanying the light chargino, which can be heavier or lighter than the chargino depending on the sign and magnitude of $M_1$. 
We must  therefore carefully consider the constraints on the lightest neutralino at LEP in our scenario with a sneutrino LSP with a displaced decay. 

The neutralino will decay to a neutrino and a sneutrino: 
\begin{equation}
\tilde \chi^0_1 \rightarrow  \nu_e + \tilde \nu_e.
\label{eq:neutralino-decay}
\end{equation}
For a displaced sneutrino decay consistent with the constraints derived in Figs.~\ref{fig:sneutrino-bound},~\ref{fig:chargino-bound}, neutralino pair production will result in the signature of neutral hadrons and missing momentum. Such a signature was never searched for by the experiments at LEP. Another possible channel we must consider is neutralino pairs produced in association with a hard photon. This can lead to events in searches for a single photon with missing energy. A number of searches were performed at LEP for this signature~\cite{Abdallah:2008aa,Abdallah:2003np,Abreu:2000vk,Achard:2003tx,Acciarri:1999kp,Abbiendi:2000hh,Abbiendi:1998yu,Barate:1997ue}. However, various cuts employed in these analyses will generally remove the signal events, {\it e.g.}, a veto on additional showers in the calorimeters~\cite{Abdallah:2003np}, which would be caused by the decay products of the sneutrino.
Finally, we also note that for small $\tan \beta$ the lightest neutralino coupling to the $Z$ boson is suppressed, and thus the pair production cross section of neutralinos will be very small at LEP. Thus, the LEP searches place no strong constraints on very light neutralinos decaying via (\ref{eq:neutralino-decay}). 

We must also consider the selectron $\tilde e_L$ that accompanies the sneutrino LSP $\tilde \nu_e$. 
In the limit that the soft masses satisfy $m^2_{\tilde E} \gg m^2_{\tilde L}$, the lightest slepton mass can be written as 
\begin{eqnarray}
m_{\tilde e_L}^2  & \simeq & m_{\tilde \nu_e}^2  - m_W^2 \cos {2\beta}.
\end{eqnarray}
Even for fairly small values of $\tan \beta$, the selectron mass can be pushed above 100 GeV and thus outside the kinematic domain of the LEP searches. As an example, for a 65 GeV electron sneutrino, the selectron mass is $m_{\tilde e_L} \approx 100$ GeV for $\tan\beta \approx 4$. 

Even if the selectron is lighter than 100 GeV, it will still be allowed by LEP searches due to the displaced decay of the sneutrino LSP. The selectron will decay via 
\begin{equation}
\tilde e_L  \rightarrow 
\begin{cases}
  \tilde \chi^-   +  \nu_e \rightarrow  e^- +  \nu_e +   \tilde \nu_e^*, \\
  \tilde \chi^0  \, +  e^- \rightarrow e^-  + \nu_e +  \tilde \nu_e^*,
\end{cases} 
\end{equation}
followed by the subsequent displaced decay of the sneutrino according to (\ref{eq:sneutrino-decay}). Slepton pair production has a final state containing two acoplanar leptons plus missing transverse momentum, and so could show up in standard R-parity conserving slepton searches. However, as discussed in the case of the chargino above, these searches employ a neutral veto which would disqualify signal events with a sneutrino displaced decay within the detector. These searches can therefore place no additional constraints on our scenario beyond those presented in Figs.~\ref{fig:sneutrino-bound},~\ref{fig:chargino-bound}.

\paragraph{SUSY searches at Tevatron and LHC}

We have demonstrated that our scenario with a very light chargino and displaced electron sneutrino LSP can not be constrained for sneutrino lifetimes $25\,{\rm cm} \lesssim c \tau_{\tilde \nu_e} \lesssim 300\,{\rm cm}$ by searches at LEP. 
We now consider additional possible direct constraints on this scenario from searches at the Tevatron and LHC including standard and RPV SUSY searches. 

The lightest states in the spectrum can be produced with substantial rates at hadron colliders. For example, the $\tilde \chi_1^+ \tilde \chi_1^-$ and $\tilde \chi_1^\pm \tilde \chi_1^0$ production cross sections range from $O(1-10\, {\rm pb})$ for charginos/neutralinos in the mass range of interest. 
Moreover, because we are considering values of $\mu$, $M_2$ of order 100 GeV, there will be additional light electroweak-inos with masses at the weak scale. Production of these particles at hadron colliders can yield via their cascade decays final states with multiple jets, leptons and missing energy. 

However, standard searches at the Tevatron and LHC looking for these signatures
place quality cuts on visible final state objects~\cite{Khachatryan:2011tk,ATLAS:2012ona,ATLAS:2011ad,ATLAS:2013rla,Abazov:2007kg,Abulencia:2006in} that are generally not suitable for the displaced decay of the neutral LSP in our scenario. 
Lepton candidates are required to have good tracks originating from primary vertex~\cite{Aad:2011mk,Aad:2010yt,CMS:pfalgo,Chatrchyan:2012xi}. 
Furthermore, jets are disqualified if the associated charged tracks carry too small a fraction of the total jet energy~\cite{Aad:2011he,ATLAS:dataquality,CMS:calojetqual}. 
But large impact parameter tracks and neutral trackless jets are expected in events with a displaced sneutrino LSP decaying to dilepton pairs in (\ref{eq:sneutrino-decay}) at the end of the cascade.
With regards to the lightest states in the spectrum, although the production rate is large the missing transverse energy is small and the prompt leptons are soft. This is simply because these states are light $\lesssim 100$ GeV and the decays feature a small $\sim {\cal O}(10\, {\rm GeV})$ mass gap. 
However, standard SUSY searches typically require large missing transverse energy and impose hard $p_T$ cuts on leptons. 
Due to these considerations, we do not expect standard SUSY searches at hadron colliders to constrain our scenario.

Despite these reservations, 
as a conservative check we have investigated the sensitivity of tri-lepton plus missing energy search~\cite{ATLAS:2013rla, CMS:aro}, which currently provides the best sensitivity for gaugino and slepton pair production. The heavier electroweak-inos in our scenario decaying via $W,Z$ bosons can yield events with multiple hard prompt leptons and neutrinos. We simulate the ATLAS search~\cite{ATLAS:2013rla} for following benchmark point (yielding an enhancement of $\sim 50\%$ in the $h\rightarrow \gamma\gamma$, $h\rightarrow \gamma Z$ decay rate):
\begin{eqnarray}
& \tan \beta = 1.5, ~ \mu = M_2  = 149\, {\rm  GeV}, ~ M_1 = 1\,{\rm TeV}, & \nonumber \\
& m_{L_1} = 76\,{\rm GeV}, ~ m_{E_1} = 1\,{\rm TeV}. & 
\end{eqnarray}
The spectrum contains a sneutrino LSP at $64$ GeV and along with a neutralino, chargino and slepton at 70, $71$ and $82$ GeV, respectively. Heavy electroweak-inos at this point have masses of $\sim 150$ GeV and 230 GeV. We generate all possible electroweak-ino pair production processes and subsequent decays with {\sc MadGraph}, interfaced with {\sc Pythia}~\cite{Sjostrand:2006za} for showering. Jets are reconstructed with the anti-$k_T$ algorithm~\cite{Cacciari:2008gp} with radius parameter $R = 0.4$ using {\sc FastJet}~\cite{Cacciari:2011ma}. We assume that the sneutrino LSP does not yield missing energy as it decays within the detector. Furthermore, we  conservatively assume that events are not rejected by any track or jet quality cuts. We find that this benchmark point is safe from the three-lepton search, which can be understood as a consequence of the smaller production rate of the heavier electroweak-inos in comparison with pure winos considered in Refs.~\cite{ATLAS:2013rla, CMS:aro}, as well as  
a suppression of missing energy due to decay of the LSP inside the detector.
We further note that it is possible to raise the mass of the heavy neutralinos/chargino as desired by splitting $\mu, M_2$, while still obtaining a sub-100 GeV chargino, though at the expense of smaller corrections to the $h\rightarrow \gamma\gamma$, $h\rightarrow \gamma Z$ decay rates. This will further weaken the sensitivity of standard SUSY searches at the LHC.

Current multi-lepton RPV searches at LHC are also not sensitive to our scenario. As with standard SUSY searches, multi-lepton RPV searches also reject events with
jets that fail to pass the quality cuts described above~\cite{ATLAS:2012153,ATLAS:2013qla,CMS:multilep} and require all leptons to come from a common primary vertex~\cite{ATLAS:2013qla,CMS:multilep,CMS:oxa}.

\paragraph{Dedicated searches for long-lived neutral particles}

A number of dedicated searches for long-lived neutral particles have been performed in the past which in principle could be sensitive to our scenario or variations thereof, which we now survey.

At LEP, such searches were primarily limited to acoplanar photons and a single non-pointing photon plus missing transverse momentum (see, {\it e.g.}, Refs.~\cite{Heister:2002vh,Barate:1999gm,Barate:1998zp}), as motivated by decays of long-lived neutralinos to a photon and gravitino in gauge mediation scenarios. 
The displaced sneutrino can potentially fake a photon if it decays near or within the electromagnetic calorimeter (ECAL). However, the muon that originates from the decay of the sneutrino will leave hits in the hadronic calorimeter (HCAL) and muon chambers. These events are thus subject to a veto designed to suppress cosmic backgrounds~\cite{Barate:1999gm}.

At the Tevatron, dedicated searches for Hidden Valley scenarios~\cite{Strassler:2006im} in events containing a displaced neutral particle decaying to two jets were performed by both CDF~\cite{Aaltonen:2011rja} 
and D0~\cite{Abazov:2009ik}. These searches do not directly apply to our scenario, but could lead to constraints on scenarios in which the sneutrino decays hadronically. D0 furthermore performed  searches for long-lived neutral particles decaying to $e^+ e^-$, $\mu^+ \mu^-$, and photon pairs~\cite{Abazov:2009ik}, \cite{Abazov:2006as}. However, these searches will not apply to displaced sneutrinos decaying to different flavor lepton pairs, such as $e\mu$ in our scenario. 

The LHC experiments are pursuing a broader program of searches sensitive to displaced decays. ATLAS has set limits on long-lived neutral particles that decay in the outer hadronic calorimeter or in the muon spectrometer (MS)~\cite{ATLAS:2012av}. The analysis employs a dedicated muon cluster RoI (Region of Interest) trigger. A muon RoI is simply a coincidence of hits in the MS. The trigger requires at least three muon RoIs in a $\Delta R = 0.4$ cone. 
The long-lived sneutrino decaying via (\ref{eq:sneutrino-decay}) will yield at most two RoIs. Therefore, the search is not applicable to our scenario.

Ref.~\cite{Aad:2012kw} performs the search for the Higgs decaying to 
lepton jets~\cite{ArkaniHamed:2008qp,Falkowski:2010cm,Falkowski:2010gv}. 
This involves at the intermediate stage long-lived neutral particles decaying to collimated muon pairs. There are several selections that make this search inapplicable to our scenario. In particular, the search reconstructs muon jets -- $\mu^+\mu^-$ pairs in a narrow cone, which are not present in our scenario. 

There is an ATLAS search for a long-lived neutralino decaying 
to a muon and two jets~\cite{Aad:2012zx}. The analysis reconstructs a displaced vertex from a muon and other charged particles. In particular, the displaced vertex is required to have at least 5 associated tracks. In our scenario, with a displaced sneutrino decaying to an $e\mu$ pair, there will only be two tracks for each displaced vertex, so this search will not apply. 

Finally, CMS has carried out searches for the Higgs decaying to two long-lived particles which subsequently decay to $e^+e^-$, $\mu^+ \mu^-$ pairs~\cite{Chatrchyan:2012jna}, and dijets pairs~\cite{CMS:dis-dijet}. 
Notably, the dilepton search does not select $e\mu$ resonances as would be present in our scenario. However, this search likely sets strong constraints on very light sneutrino LSP with a displaced decay to $e^+e^-$ or $\mu^+ \mu^-$. The dijet search imposes a number of cuts which are not well-suited for our scenario, {\it e.g.}, hard jet $p_T$ and scalar $p_T$ sum $H_T$ cuts, as well as minimum vertex multiplicity selections.

\paragraph{\bf Testing the scenario at the LHC}

We now discuss how this scenario can be tested at LHC. The characteristic signature is a displaced $e\mu$ resonance arising when a sneutrino decays in the inner detector.
The experimental techniques necessary to search for this signature can likely be adapted from existing searches for prompt  different-flavor dilepton resonances~\cite{Aad:2012bwa} and displaced same-flavor dilepton resonances~\cite{Chatrchyan:2012jna}.  
For example, the displaced same-flavor search~\cite{Chatrchyan:2012jna} requires either two energy deposits in the ECAL
or two track segments in MS, along with associated displaced
tracks without track quality criteria imposed.
Similarly, the signal events in our scenario can be selected by one ECAL deposit + one MS track.
Although cosmic muons can fake a single MS track trigger, they can be suppressed by the further requirement of an associated displaced vertex pointing back to primary vertex.
This search maybe useful for sneutrino decays within about 50 cm~\cite{Chatrchyan:2012jna,CMS:tracking} where tracking is available. 

If one or both of the sneutrinos in our scenario decay near the ECAL and consequently some tracks are not reconstructable, an alternative trigger based on non-standard objects such as trackless jets (objects characterized by calorimeter hits isolated from tracks in the inner detector) can be employed. ATLAS has developed dedicated triggers for trackless jets utilizing either 1) an associated trackless muon, or 2) large HCAL to ECAL energy deposition~\cite{Aad:2013txa}. 
Finally, we note that the displaced vertex reconstruction techniques utilizing capabilities of different parts of detectors (not just of tracker) discussed in Ref.~\cite{Meade:2010ji} may also be useful for our scenario.

\paragraph{\bf Other scenarios for light charginos}

Here we wish to make some preliminary comments on variations of the basic scenario considered above, reserving a detailed investigation for future work. 
If the sneutrino has a different decay mode (\ref{eq:sneutrino-decay}), there can be additional constraints which are applicable. For instance, 
if the sneutrino decays to a dielectron or dimuon pair, then the CMS search for two displaced $e^+e^-$ or $\mu^+ \mu^-$ resonances~\cite{Chatrchyan:2012jna} will place stronger constraints on the lower end of the sneutrino lifetime range than the LEP RPV searches displayed in Fig.~\ref{fig:sneutrino-bound}. This highlights the power of dedicated searches for displaced neutral particles. 

However, final states containing displaced resonances with $\tau$ leptons are not explicitly covered by the CMS displaced same-flavor dilepton search~\cite{Chatrchyan:2012jna}. In particular, the search requires that the reconstructed dilepton resonance momentum vector is collinear with the vector pointing from the primary vertex to the
displaced vertex, which will generally not be satisfied as the $\tau$ shares its energy among the final state lepton and neutrinos. Thus, sneutrinos decaying at a macroscopic distance to $e\tau$, $\mu\tau$, or $\tau\tau$ may also still be viable. Dedicated searches for these modes should be performed.

If the sneutrino decays to $q \bar q$ through a $LQD$ operator, then the situation is somewhat more complicated. 
RPV $UDD$ searches at LEP involving final state jets generally require 8 or more good charged tracks 
({\it i.e.},  tracks with small impact parameter). 
Despite the fact that the sneutrino is displaced, an appreciable efficiency to meet the track multiplicity requirements is expected even for lifetimes $\gtrsim 10$ cm because each quark can yield $O(10)$ charged particles during hadronization. Preliminary simulations confirm this expectation, although there is likely still an open window for this scenario from LEP searches for sneutrino lifetimes of order 100 cm. At the LHC, searches for displaced vertices formed purely from jets are difficult due to the multijet background. 
Either an extra muon is required~\cite{Aad:2012zx,Aad:2013txa}, or dedicated vertex track requirements are utilized~\cite{CMS:dis-dijet}. 
The latter search, which likely has some sensitivity to this scenario, imposes strong $p_T$ and $H_T$ cuts which are not ideal for very light, long-lived sneutrinos decaying to dijet pairs. A search targeted at low-mass, long-lived dijet resonances should be developed.

Another scenario deserving of consideration contains a chargino NLSP and neutralino LSP with a displaced RPV decay (sneutrino and slepton are heavy). In this scenario, the chargino decays via $\tilde \chi^+_1 \rightarrow \tilde \chi^0 W^{+^*}$. The prompt hadronic decays of the virtual $W^{+*}$ boson will lead to many good charged tracks, implying that the track multiplicity selection will be highly efficient. It therefore appears difficult to hide a light chargino in this scenario.

\section{Precision electroweak data}

As we have demonstrated above, the very light charginos in our scenario can generate sizable one-loop corrections to the effective couplings $h\gamma\gamma$ and $h\gamma Z$. One might therefore expect similarly large one-loop contributions to the gauge boson vacuum polarizations, which can affect the predictions for precision electroweak observables. To investigate this issue, we have performed a global fit to the precision data along the lines of the Gfitter group~\cite{Baak:2012kk}. A detailed description of the experimental observables and theoretical predictions entering in the fit can be found in Ref.~\cite{Batell:2012ca}.

The modifications to the $W,Z,\gamma$ vacuum polarizations are encoded in the $STUVWX$ extended oblique parameters of Ref.~\cite{Maksymyk:1993zm,Burgess:1993mg}. This extended formalism is necessary since we are considering charginos masses of order the $Z$ boson mass and lighter. In the Appendix we report the predictions for the observables entering into our fit in terms of these oblique parameters. For the chargino and neutralino contributions to the gauge boson vacuum polarizations we use the results of Ref.~\cite{Martin:2004id}.

\begin{figure}
\centerline{
\includegraphics[width= 0.93\columnwidth]{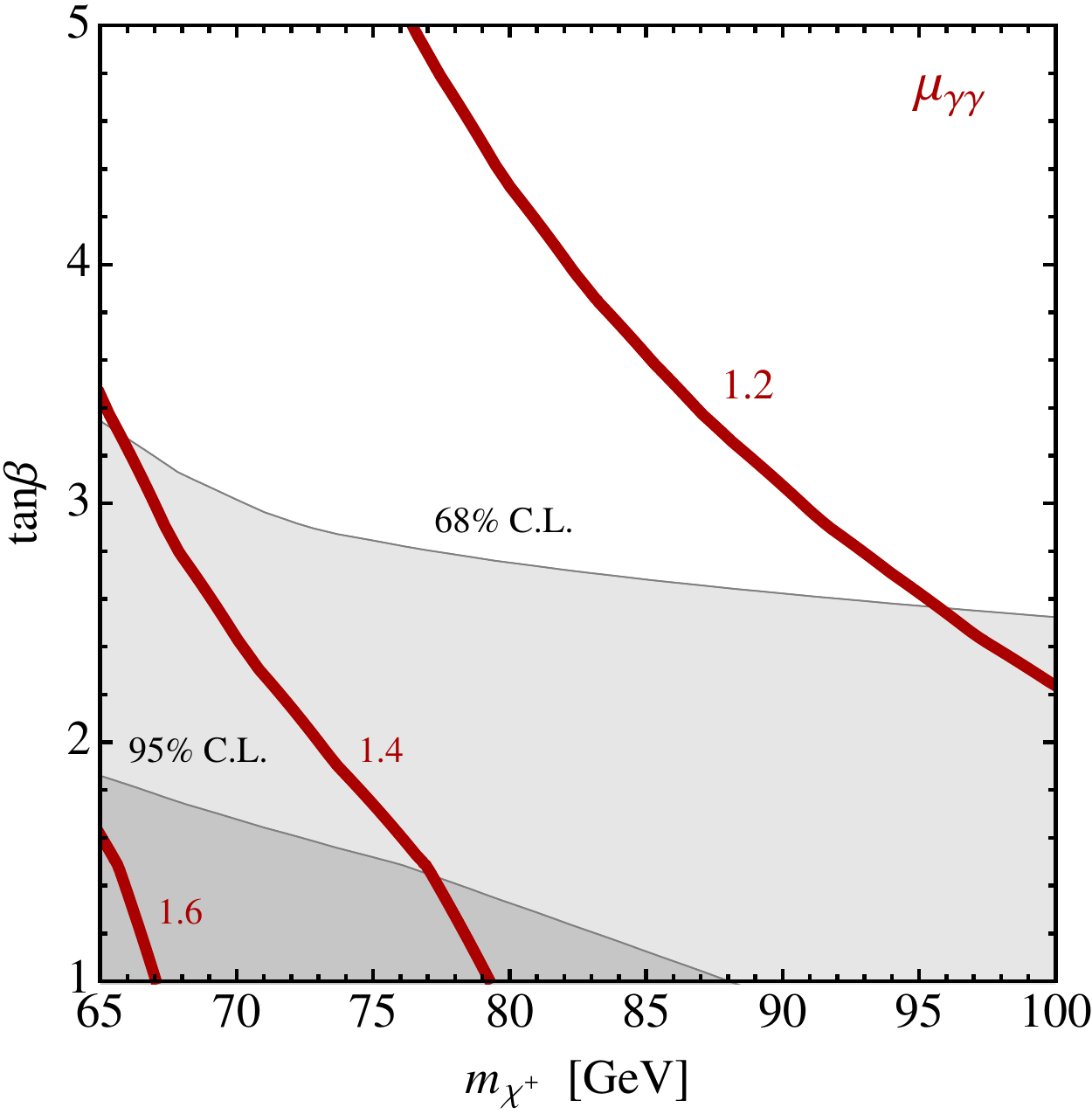}
}
\caption{Precision electroweak data $68\%$, $95\%$ C.L. contours (black) in the $m_{\tilde \chi_1^+} - \tan\beta$ plane. We also display in red contours of the chargino contribution to the diphoton signal strength. Here we have fixed $M_2 = \mu$, $M_1 = 1$ TeV. }
\label{fig:EWPD1}
\end{figure}

In Figure.~\ref{fig:EWPD1} we display in black the $68\%$, $95\%$ C.L. contours from our fit to the precision data for the case of $M_2 = \mu$, $M_1 = 1$ TeV  in the chargino mass - $\tan \beta$ plane. For comparison we have also overlaid in red contours of the chargino contribution to the diphoton signal strength. The best fit in this region is obtained for $\tan\beta \approx 5$ and a chargino mass around 80 GeV, although the $\chi^2$ is essentially flat in $m_{\tilde \chi^+_1}$ above this value. The total $\chi^2$ at the best fit point, $\chi^2_{\rm min} = 19.7$ is somewhat smaller than the SM value, $\chi^2_{\rm SM} = 20.7$. 

We observe in Figure.~\ref{fig:EWPD1} that the theoretical description of the precision data becomes worse for lighter charginos and small values of $\tan \beta$. There are two observables in this region (beyond the already discrepant $A_{FB}^b$ and $R_b$) that display a slight $\sim 2 \sigma$ tension: 1) The $W$ boson mass, which the SM predicts to be $(m_W)_{\rm SM} = 80.362 \pm 0.007$ GeV, becomes smaller as a result of a small negative $U$ parameter, widening the gap with the experimental value $(m_W)_{\rm exp} = 80.385 \pm 0.015$ GeV. 2) The leptonic asymmetry parameter $A_\ell$, predicted in the SM to be  $(A_\ell)_{\rm SM} = 0.1472 \pm 0.0007$, becomes smaller due to a small positive $X$ parameter, increasing the tension with the experimental average $(A_\ell)_{\rm exp} = 0.1499 \pm 0.0018$. 

The tension in this region can be relaxed with a small positive $T$ parameter from a source other than the charginos.
A new contribution of order $T \sim 0.05 - 0.1$ raises the values of $m_W$ and $A_\ell$, 
leading to a global fit that resembles that of the SM.  
As an example, a light, highly-mixed stop 
can easily yield the required contribution to $T$, while at the same time not upsetting the observed Higgs production rates. This is because the 
coupling of the light stops to the Higgs is proportional to $\sim (1 - A_t^2/m_{U_3}^2)$ and therefore only mildly disturbs the gluon fusion rate if $A_t \sim m_{U_3}$.
For instance, with the inputs $m_{Q_3} = 200$ GeV, $m_{U_3} = 1.7$ TeV, $A_t = 1.2$ TeV, $\mu = 150$ GeV, $\tan \beta = 1.5$,
one obtains a stop with mass close to the top, a contribution $T \sim 0.07$, as well as a small $10\%$ enhancement of the gluon fusion rate. See the discussion in Ref.~\cite{Carena:2013iba} for more details. 
A detailed investigation of the collider bounds on the light stop and sbottom in this case is beyond the scope of this paper, although the presence of a displaced sneutrino LSP in the decay chain implies that these events 1) contain significantly less missing energy than in standard R-parity conserving scenarios and 2) may be subject to the jet and charged track quality cuts described above. 

\section{Discussion and outlook}

In this work we have investigated a scenario with a very light chargino in the mass range $m_h/2 < m_{\tilde \chi^+} \lesssim 100$ GeV, below the naive kinematic reach of LEP2.  A chargino in this mass range can lead to dramatic deviations in the $h\rightarrow \gamma \gamma$ and  $h\rightarrow Z \gamma$ decay rates, which could be measured at the LHC and future high energy colliders. A variety of LEP2, Tevatron, and LHC searches place strong constraints on this hypothesis, but the analyses are model dependent and based on assumed decay channels of the charginos. We presented a concrete scenario which is not covered by existing searches: a chargino decaying to sneutrino LSP, which subsequently decays through a small RPV coupling with a macroscopic decay length of order $10\,{\rm cm} - 100\,{\rm cm}$. The charged particles in the sneutrino decay products generically fail to point back to the primary vertex, and as such do not pass basic track impact parameter selection cuts required at LEP. Furthermore, standard SUSY searches at LEP looking for missing transverse momentum are weakened by a neutral veto which will be in effect unless both sneutrinos decay outside the detector.

The 125 GeV Higgs mass can be obtained in this scenario in several ways. In the region of low $\tan\beta$ relevant for modifications to the $h\rightarrow \gamma\gamma, Z \gamma$ rate, it is difficult to obtain a Higgs mass of 125 GeV in the MSSM with superpartners at the TeV scale. One possibility is to simply raise the masses of the scalars (except for the light sneutrino and slepton), and in particular the stops into the multi-TeV range. Of course, such heavy stops require a finely-tuned weak scale, and in this scenario one cannot rely on the stops to improve the precision electroweak data.  For slightly larger values $\tan\beta \gtrsim 4$ it is also possible to obtain the observed Higgs mass in the MSSM with highly mixed stops at the TeV scale. Finally, one can investigate extensions of the MSSM with new $F$- or $D$-term contributions to the Higgs mass, see for example Refs.~\cite{Kane:1992kq,Espinosa:1992hp,Ellwanger:2009dp,Batra:2003nj,Lu:2013cta}. 
These extensions can lead to modifications of the chargino interactions with the Higgs (see, {\it e.g.}, Refs.~\cite{Huo:2012tw},\cite{Gherghetta:2012gb}).

While our discussion has focused on charginos, similar considerations apply to other hypothetical light charged particles. With regards to the corrections to $h\rightarrow \gamma\gamma$, the one-loop form factor for scalars, fermions, as vectors exhibit a rise in magnitude as the mass of the charged particle in the loop approaches $m_h/2$, as we have illustrated for the case of the fermion in Fig.~\ref{fig:form-factor}. The basic scenario for a hidden chargino described in this work can, with a few modifications, be used to hide other light charged particles. For example, a light stau NLSP decaying to 
a long-lived neutralino LSP will be weakly constrained if the neutralino lifetime is  $O(10\,{\rm cm} - 100\,{\rm cm})$. It would also be interesting to consider non-supersymmetric scenarios. For example, light vector-like leptons above ~100 GeV can lead to large deviations in $h\rightarrow \gamma\gamma$ at the expense of a low scale $\sim 10$ TeV vacuum instability
~\cite{Joglekar:2012vc,ArkaniHamed:2012kq,Reece:2012gi,Batell:2012mj,Kearney:2012zi,McKeen:2012av,Davoudiasl:2012ig,
Batell:2012zw,Fan:2013qn,Joglekar:2013zya,Altmannshofer:2013zba}.
However, if the charged states are lighter are below 100 GeV, the same effect can be obtained with much weaker couplings of the vector-like leptons to the Higgs, allowing for a much higher UV completion scale. 

Given this window in which light charginos, and by extension various other hypothetical light charged particles with displaced neutral particles in their decay products, are unconstrained, it is important that the LHC experiments develop dedicated searches to probe this scenario. We have touched on a few possible strategies to cover the scenario proposed here, although detailed feasibility studies and concrete search strategies are still required. Furthermore, it should be possible to reanalyze the LEP2 data to test this scenario.  While the current motivation for such an effort is simply to cover an interesting open window in SUSY parameter space, 
we stress that any future definitive observation of a discrepancy in the $h\rightarrow \gamma\gamma$ rate will 
mandate reconsideration of the possibility of charged particles below 100 GeV.

\subsubsection*{\bf Acknowledgements}
We thank Y.~Gershtein, P.~Ko, H.~M.~Lee, S.~Martin, T. Roy, P.~Saraswat, and L.~T.~Wang for helpful discussions.  
Work at ANL is supported in part by the U.S. Department of Energy under Contract No. DE-AC02-06CH11357. 
B.B. is supported by the NSF under grant PHY-0756966 and the DOE Early Career Award under grant DE-SC0003930.
S.J. thanks KIAS Center for Advanced Computation for providing computing resources. 
B.B. and C.W. thank the Aspen Center for Physics and the KITP, Santa Barbara, where part of the work has been done. 
B.B. also thanks KIAS and the 2013 Santa Fe workshop INFO, sponsored by Los Alamos National Laboratory, where part of this work was completed.

\appendix*
\def\theequation{\thesection A.\arabic{equation}}
\setcounter{equation}{0}
\section{Appendix~~~Predictions for Electroweak Observables}
\label{ap:beta}

The predictions for the precision observables entering into our fit in terms of $STUVWX$
 can be obtained from Refs.~\cite{Maksymyk:1993zm,Burgess:1993mg} and are as follows:
\begin{eqnarray}
 \Gamma_Z & = & 2.4957 -0.0092 S+0.026 T+0.019 V-0.020 X,~~      \nonumber \\
 \sigma^0_{\rm had} & = & 41.475 +0.014 S-0.0098 T+0.031 X,    \nonumber \\           
 R_\ell & = &  20.744 -0.062 S+0.042 T-0.14 X,  \nonumber \\
 A_{FB}^{\ell} & = &0.01625 -0.0061 S+0.0042 T-0.013 X,  \nonumber \\
 A_\ell & = & 0.1472 -0.028 S+0.019 T-0.061 X,  \nonumber \\      
 A_c & = & 0.6680 -0.012 S+0.0084 T-0.027 X,  \nonumber \\
 A_b & = & 0.93468 -0.0023 S+0.0016 T-0.0050 X,  \nonumber \\                            
 A_{FB}^c & = & 0.0738 -0.015 S+0.010 T-0.033 X,    \\         
 A_{FB}^b & = & 0.1032 -0.020 S+0.014 T-0.043 X,     \nonumber \\
 R_c & = & 0.1723 -0.00021 S+0.00015 T -0.00046 X,  \nonumber \\
 R_b & = & 0.21475 +0.00013 S-0.000091 T+0.00030 X,       \nonumber \\
 s^2_{\theta_{\rm  eff}} & = &  0.23149 +0.0035 S-0.0024 T+0.0078 X,      \nonumber \\
 m_W & = &  80.362 -0.28 S+0.43 T+0.35 U, \nonumber \\
 \Gamma_W & = & 2.091 -0.015 S+0.023 T+0.018 U+0.016 W.   \nonumber 
\end{eqnarray}
For the results above, we have fixed the SM inputs $(m_Z, m_t, m_h, \alpha_s, \Delta \alpha^{(5)}_{\rm had})$ to the predicted values of Gfitter obtained in Ref.~\cite{Baak:2012kk}.


\begin{thebibliography}{99}


\bibitem{ATLAS-Higgs} 
  G.~Aad {\it et al.}  [ATLAS Collaboration],
  Phys.\ Lett.\ B
  [arXiv:1207.7214 [hep-ex]].

\bibitem{CMS-Higgs}
  S.~Chatrchyan {\it et al.}  [CMS Collaboration],
  Phys.\ Lett.\ B
  [arXiv:1207.7235 [hep-ex]].
  

\bibitem{Casas:2013pta} 
  J.~A.~Casas, J.~M.~Moreno, K.~Rolbiecki and B.~Zaldivar,
  arXiv:1305.3274 [hep-ph].


\bibitem{Graham:2012th} 
  P.~W.~Graham, D.~E.~Kaplan, S.~Rajendran and P.~Saraswat,
  JHEP {\bf 1207}, 149 (2012)
  [arXiv:1204.6038 [hep-ph]].

  
\bibitem{Carena:2011aa} 
  M.~Carena, S.~Gori, N.~R.~Shah and C.~E.~M.~Wagner,
  JHEP {\bf 1203}, 014 (2012)
  [arXiv:1112.3336 [hep-ph]].
  
\bibitem{Carena:2012gp} 
  M.~Carena, S.~Gori, N.~R.~Shah, C.~E.~M.~Wagner and L.~-T.~Wang,
  JHEP {\bf 1207}, 175 (2012)
  [arXiv:1205.5842 [hep-ph]].
  
\bibitem{Carena:2012mw} 
  M.~Carena, S.~Gori, I.~Low, N.~R.~Shah and C.~E.~M.~Wagner,
  JHEP {\bf 1302}, 114 (2013)
  [arXiv:1211.6136 [hep-ph]].
  
\bibitem{Carena:2013iba} 
  M.~Carena, S.~Gori, N.~R.~Shah, C.~E.~M.~Wagner and L.~-T.~Wang,
  arXiv:1303.4414 [hep-ph].


\bibitem{ATLAS-LP}
J.~Vossebeld (on behalf of the ATLAS Collaboration), talk at 
{\it 21st International Conference on Supersymmetry and Unification of Fundamental Interactions}, 
August 26- 31, 2013, ICTP Trieste, Italy

\bibitem{CMS-LP}
S.~Dasu (on behalf of the CMS Collaboration), talk at 
{\it 21st International Conference on Supersymmetry and Unification of Fundamental Interactions}, 
August 26- 31, 2013, ICTP Trieste, Italy
   

\bibitem{Kalyniak:1985ct} 
  P.~Kalyniak, R.~Bates and J.~N.~Ng,
  Phys.\ Rev.\ D {\bf 33}, 755 (1986).

\bibitem{Bates:1986zv} 
  R.~Bates, J.~N.~Ng and P.~Kalyniak,
  Phys.\ Rev.\ D {\bf 34}, 172 (1986).
  
\bibitem{Gunion:1988mf} 
  J.~F.~Gunion, G.~Gamberini and S.~F.~Novaes,
  Phys.\ Rev.\ D {\bf 38}, 3481 (1988).

\bibitem{Djouadi:1996pb} 
  A.~Djouadi, V.~Driesen, W.~Hollik and J.~I.~Illana,
  Eur.\ Phys.\ J.\ C {\bf 1}, 149 (1998)
  [hep-ph/9612362].
  
\bibitem{Weiler:1988xn} 
  T.~J.~Weiler and T.~-C.~Yuan,
  Nucl.\ Phys.\ B {\bf 318}, 337 (1989).
 
\bibitem{Djouadi:1996yq} 
  A.~Djouadi, V.~Driesen, W.~Hollik and A.~Kraft,
  Eur.\ Phys.\ J.\ C {\bf 1}, 163 (1998)
  [hep-ph/9701342].
  
\bibitem{Lee:2012wa} 
  J.~S.~Lee, M.~Carena, J.~Ellis, A.~Pilaftsis and C.~E.~M.~Wagner,
  Comput.\ Phys.\ Commun.\  {\bf 184}, 1220 (2013)
  [arXiv:1208.2212 [hep-ph]].
  
\bibitem{Lee:2007gn} 
  J.~S.~Lee, M.~Carena, J.~Ellis, A.~Pilaftsis and C.~E.~M.~Wagner,
  Comput.\ Phys.\ Commun.\  {\bf 180}, 312 (2009)
  [arXiv:0712.2360 [hep-ph]].
 
  
\bibitem{Ellis:1975ap} 
  J.~R.~Ellis, M.~K.~Gaillard and D.~V.~Nanopoulos,
  Nucl.\ Phys.\ B {\bf 106}, 292 (1976).
 
\bibitem{Shifman:1979eb} 
  M.~A.~Shifman, A.~I.~Vainshtein, M.~B.~Voloshin and V.~I.~Zakharov,
  Sov.\ J.\ Nucl.\ Phys.\  {\bf 30}, 711 (1979)
  [Yad.\ Fiz.\  {\bf 30}, 1368 (1979)].


\bibitem{Djouadi:2005gi} 
  A.~Djouadi,
  Phys.\ Rept.\  {\bf 457}, 1 (2008)
  [hep-ph/0503172].
  

\bibitem{Peskin:2012we} 
  M.~E.~Peskin,
  arXiv:1207.2516 [hep-ph].
  
\bibitem{Klute:2013cx} 
  M.~Klute, R.~Lafaye, T.~Plehn, M.~Rauch and D.~Zerwas,
  Europhys.\ Lett.\  {\bf 101}, 51001 (2013)
  [arXiv:1301.1322 [hep-ph]].
  
\bibitem{Gainer:2011aa} 
  J.~S.~Gainer, W.~-Y.~Keung, I.~Low and P.~Schwaller,
  Phys.\ Rev.\ D {\bf 86}, 033010 (2012)
  [arXiv:1112.1405 [hep-ph]].
  
\bibitem{Campbell:2013hz} 
  J.~M.~Campbell, R.~K.~Ellis, W.~T.~Giele and C.~Williams,
  Phys.\ Rev.\ D {\bf 87}, 073005 (2013)
  [arXiv:1301.7086 [hep-ph]].
  
  
\bibitem{Kribs:2008hq} 
  G.~D.~Kribs, A.~Martin and T.~S.~Roy,
  JHEP {\bf 0901}, 023 (2009)
  [arXiv:0807.4936 [hep-ph]].
  

\bibitem{Abdallah:2002rd} 
  J.~Abdallah {\it et al.}  [DELPHI Collaboration],
  Eur.\ Phys.\ J.\ C {\bf 27}, 153 (2003)
  [hep-ex/0303025].

\bibitem{Abbiendi:2003yd} 
  G.~Abbiendi {\it et al.}  [OPAL Collaboration],
  Phys.\ Lett.\ B {\bf 572}, 8 (2003)
  [hep-ex/0305031].


\bibitem{Yamagata:1993jq} 
  T.~Yamagata, Y.~Takamori and H.~Utsunomiya,
  Phys.\ Rev.\ D {\bf 47}, 1231 (1993).


\bibitem{neutralino-chargino-RPC}
LEP2 SUSY Working Group,
{\it Combined LEP Chargino Results, up to 208 GeV
for large m0},
ALEPH, DELPHI, L3, OPAL Experiments
\url{http://lepsusy.web.cern.ch/lepsusy/www/inos_moriond01/charginos_pub.html}

\bibitem{neutralino-chargino-RPC-small-DM}
LEP2 SUSY Working Group,
{\it Combined LEP Chargino Results, up to 208 GeV for low DM},
ALEPH, DELPHI, L3, OPAL Experiments
\url{http://lepsusy.web.cern.ch/lepsusy/www/inoslowdmsummer02/charginolowdm_pub.html}


\bibitem{slepton}
LEP2 SUSY Working Group,
{\it Combined LEP Selectron/Smuon/Stau Results, 183-208 GeV}
ALEPH, DELPHI, L3, OPAL Experiments
\url{http://lepsusy.web.cern.ch/lepsusy/www/sleptons_summer04/slep_final.html}
  
\bibitem{Heister:2001nk} 
  A.~Heister {\it et al.}  [ALEPH Collaboration],
  Phys.\ Lett.\ B {\bf 526}, 206 (2002)
  [hep-ex/0112011].


\bibitem{Heister:2002jc} 
  A.~Heister {\it et al.}  [ALEPH Collaboration],
  Eur.\ Phys.\ J.\ C {\bf 31}, 1 (2003)
  [hep-ex/0210014].

\bibitem{Abdallah:2003xc} 
  J.~Abdallah {\it et al.}  [DELPHI Collaboration],
  Eur.\ Phys.\ J.\ C {\bf 36}, 1 (2004)
  [Eur.\ Phys.\ J.\ C {\bf 37}, 129 (2004)]
  [hep-ex/0406009].

\bibitem{Abbiendi:2003rn} 
  G.~Abbiendi {\it et al.}  [OPAL Collaboration],
  Eur.\ Phys.\ J.\ C {\bf 33}, 149 (2004)
  [hep-ex/0310054].
   
\bibitem{Achard:2001ek} 
  P.~Achard {\it et al.}  [L3 Collaboration],
  Phys.\ Lett.\ B {\bf 524}, 65 (2002)
  [hep-ex/0110057].


\bibitem{Buskulic:1994wz} 
  D.~Buskulic {\it et al.}  [ALEPH Collaboration],
  Nucl.\ Instrum.\ Meth.\ A {\bf 360}, 481 (1995).


\bibitem{Barbier:2004ez} 
  R.~Barbier, C.~Berat, M.~Besancon, M.~Chemtob, A.~Deandrea, E.~Dudas, P.~Fayet and S.~Lavignac {\it et al.},
  Phys.\ Rept.\  {\bf 420}, 1 (2005)
  [hep-ph/0406039].


\bibitem{madgraph}
  J.~Alwall, M.~Herquet, F.~Maltoni, O.~Mattelaer and T.~Stelzer,
  JHEP {\bf 1106} (2011) 128.

\bibitem{Christensen:2008py} 
  N.~D.~Christensen and C.~Duhr,
  Comput.\ Phys.\ Commun.\  {\bf 180}, 1614 (2009)
  [arXiv:0806.4194 [hep-ph]].


\bibitem{Barate:1997ra} 
  R.~Barate {\it et al.}  [ALEPH Collaboration],
  Eur.\ Phys.\ J.\ C {\bf 4}, 433 (1998)
  [hep-ex/9712013].


\bibitem{Dreiner:2008tw} 
  H.~K.~Dreiner, H.~E.~Haber and S.~P.~Martin,
  Phys.\ Rept.\  {\bf 494}, 1 (2010)
  [arXiv:0812.1594 [hep-ph]].


\bibitem{Abdallah:2003xe} 
  J.~Abdallah {\it et al.}  [DELPHI Collaboration],
  Eur.\ Phys.\ J.\ C {\bf 31}, 421 (2003)
  [hep-ex/0311019].

\bibitem{Achard:2003ge} 
  P.~Achard {\it et al.}  [L3 Collaboration],
  Phys.\ Lett.\ B {\bf 580}, 37 (2004)
  [hep-ex/0310007].

\bibitem{Abbiendi:2003ji} 
  G.~Abbiendi {\it et al.}  [OPAL Collaboration],
  Eur.\ Phys.\ J.\ C {\bf 32}, 453 (2004)
  [hep-ex/0309014].


\bibitem{Fan:2012jf} 
  J.~Fan, M.~Reece and J.~T.~Ruderman,
  JHEP {\bf 1207}, 196 (2012)
  [arXiv:1201.4875 [hep-ph]].


\bibitem{Abdallah:2008aa} 
  J.~Abdallah {\it et al.}  [DELPHI Collaboration],
  Eur.\ Phys.\ J.\ C {\bf 60}, 17 (2009)
  [arXiv:0901.4486 [hep-ex]].

\bibitem{Abdallah:2003np} 
  J.~Abdallah {\it et al.}  [DELPHI Collaboration],
  Eur.\ Phys.\ J.\ C {\bf 38}, 395 (2005)
  [hep-ex/0406019].

\bibitem{Abreu:2000vk} 
  P.~Abreu {\it et al.}  [DELPHI Collaboration],
  Eur.\ Phys.\ J.\ C {\bf 17}, 53 (2000)
  [hep-ex/0103044].

\bibitem{Achard:2003tx} 
  P.~Achard {\it et al.}  [L3 Collaboration],
  Phys.\ Lett.\ B {\bf 587}, 16 (2004)
  [hep-ex/0402002].

\bibitem{Acciarri:1999kp} 
  M.~Acciarri {\it et al.}  [L3 Collaboration],
  Phys.\ Lett.\ B {\bf 470}, 268 (1999)
  [hep-ex/9910009].

\bibitem{Abbiendi:2000hh} 
  G.~Abbiendi {\it et al.}  [OPAL Collaboration],
  Eur.\ Phys.\ J.\ C {\bf 18}, 253 (2000)
  [hep-ex/0005002].

\bibitem{Abbiendi:1998yu} 
  G.~Abbiendi {\it et al.}  [OPAL Collaboration],
  Eur.\ Phys.\ J.\ C {\bf 8}, 23 (1999)
  [hep-ex/9810021].

\bibitem{Barate:1997ue} 
  R.~Barate {\it et al.}  [ALEPH Collaboration],
  Phys.\ Lett.\ B {\bf 420}, 127 (1998)
  [hep-ex/9710009].


\bibitem{Khachatryan:2011tk} 
  V.~Khachatryan {\it et al.}  [CMS Collaboration],
  Phys.\ Lett.\ B {\bf 698}, 196 (2011)
  [arXiv:1101.1628 [hep-ex]].

\bibitem{ATLAS:2012ona} 
  [ATLAS Collaboration],
  ATLAS-CONF-2012-109.

\bibitem{ATLAS:2011ad} 
  G.~Aad {\it et al.}  [ATLAS Collaboration],
  Phys.\ Rev.\ D {\bf 85}, 012006 (2012)
  [arXiv:1109.6606 [hep-ex]].

\bibitem{ATLAS:2013rla} 
  [ATLAS Collaboration],
  ATLAS-CONF-2013-035.

\bibitem{Abazov:2007kg} 
  V.~M.~Abazov {\it et al.}  [D0 Collaboration],
  Phys.\ Rev.\ D {\bf 76}, 092007 (2007)
  [arXiv:0705.2788 [hep-ex]].
  
\bibitem{Abulencia:2006in} 
  A.~Abulencia {\it et al.}  [CDF Collaboration],
  Phys.\ Rev.\ Lett.\  {\bf 97}, 082004 (2006)
  [hep-ex/0606017].
  
\bibitem{Aad:2011mk} 
  G.~Aad {\it et al.}  [ATLAS Collaboration],
  Eur.\ Phys.\ J.\ C {\bf 72}, 1909 (2012)
  [arXiv:1110.3174 [hep-ex]].

\bibitem{Aad:2010yt}
  G.~Aad {\it et al.}  [Atlas Collaboration],
  JHEP {\bf 1012}, 060 (2010)
  [arXiv:1010.2130 [hep-ex]].

\bibitem{CMS:pfalgo} 
  [CMS Collaboration],
  CMS-PAS-PFT-09-001

\bibitem{Chatrchyan:2012xi} 
  S.~Chatrchyan {\it et al.}  [CMS Collaboration],
  JINST {\bf 7}, P10002 (2012)
  [arXiv:1206.4071 [physics.ins-det]].

    
\bibitem{Aad:2011he}
  G.~Aad {\it et al.}  [ATLAS Collaboration],
  arXiv:1112.6426 [hep-ex].
    
\bibitem{ATLAS:dataquality}
  [ATLAS Collaboration],
  ATLAS-CONF-2010-038.

\bibitem{CMS:calojetqual} 
  [CMS Collaboration],
  CMS-PAS-JME-09-008.

 
\bibitem{CMS:aro} 
  [CMS Collaboration],
  CMS-PAS-SUS-13-006.

  
\bibitem{Sjostrand:2006za} 
  T.~Sjostrand, S.~Mrenna and P.~Z.~Skands,
  JHEP {\bf 0605}, 026 (2006)
  [hep-ph/0603175].
  

\bibitem{Cacciari:2008gp} 
  M.~Cacciari, G.~P.~Salam and G.~Soyez,
  JHEP {\bf 0804}, 063 (2008)
  [arXiv:0802.1189 [hep-ph]].

\bibitem{Cacciari:2011ma} 
  M.~Cacciari, G.~P.~Salam and G.~Soyez,
  Eur.\ Phys.\ J.\ C {\bf 72}, 1896 (2012)
  [arXiv:1111.6097 [hep-ph]].


\bibitem{ATLAS:2012153} 
  G.~Aad {\it et al.}  [ATLAS Collaboration],
  ATLAS-CONF-2012-153

\bibitem{ATLAS:2013qla} 
  [ATLAS Collaboration],
  ATLAS-CONF-2013-036.

\bibitem{CMS:multilep}
[CMS Collaboration],
CMS PAS SUS-13-010

\bibitem{CMS:oxa} 
  [CMS Collaboration],
  CMS-PAS-SUS-12-026.


\bibitem{Heister:2002vh} 
  A.~Heister {\it et al.}  [ALEPH Collaboration],
  Eur.\ Phys.\ J.\ C {\bf 25}, 339 (2002)
  [hep-ex/0203024].

\bibitem{Barate:1999gm} 
  R.~Barate {\it et al.}  [ALEPH Collaboration],
  Eur.\ Phys.\ J.\ C {\bf 16}, 71 (2000).

\bibitem{Barate:1998zp} 
  R.~Barate {\it et al.}  [ALEPH Collaboration],
  Phys.\ Lett.\ B {\bf 433}, 176 (1998).

\bibitem{Strassler:2006im} 
  M.~J.~Strassler and K.~M.~Zurek,
  Phys.\ Lett.\ B {\bf 651}, 374 (2007)
  [hep-ph/0604261].

\bibitem{Aaltonen:2011rja} 
  T.~Aaltonen {\it et al.}  [CDF Collaboration],
  Phys.\ Rev.\ D {\bf 85}, 012007 (2012)
  [arXiv:1109.3136 [hep-ex]].
  
\bibitem{Abazov:2009ik} 
  V.~M.~Abazov {\it et al.}  [D0 Collaboration],
  Phys.\ Rev.\ Lett.\  {\bf 103}, 071801 (2009)
  [arXiv:0906.1787 [hep-ex]].

\bibitem{Abazov:2006as} 
  V.~M.~Abazov {\it et al.}  [D0 Collaboration],
  Phys.\ Rev.\ Lett.\  {\bf 97}, 161802 (2006)
  [hep-ex/0607028].
  
\bibitem{ATLAS:2012av} 
  G.~Aad {\it et al.}  [ATLAS Collaboration],
  Phys.\ Rev.\ Lett.\  {\bf 108}, 251801 (2012)
  [arXiv:1203.1303 [hep-ex]].

\bibitem{Aad:2012kw} 
  G.~Aad {\it et al.}  [ATLAS Collaboration],
  Phys.\ Lett.\ B {\bf 721}, 32 (2013)
  [arXiv:1210.0435 [hep-ex]].

\bibitem{ArkaniHamed:2008qp} 
  N.~Arkani-Hamed and N.~Weiner,
  JHEP {\bf 0812}, 104 (2008)
  [arXiv:0810.0714 [hep-ph]].

\bibitem{Falkowski:2010cm} 
  A.~Falkowski, J.~T.~Ruderman, T.~Volansky and J.~Zupan,
  JHEP {\bf 1005}, 077 (2010)
  [arXiv:1002.2952 [hep-ph]].

\bibitem{Falkowski:2010gv}
  A.~Falkowski, J.~T.~Ruderman, T.~Volansky and J.~Zupan,
  Phys.\ Rev.\ Lett.\  {\bf 105} (2010) 241801
  [arXiv:1007.3496 [hep-ph]].

\bibitem{Aad:2012zx} 
  G.~Aad {\it et al.}  [ATLAS Collaboration],
  Phys.\ Lett.\ B {\bf 719}, 280 (2013)
  [arXiv:1210.7451 [hep-ex]].
  
\bibitem{Chatrchyan:2012jna} 
  S.~Chatrchyan {\it et al.}  [CMS Collaboration],
  JHEP {\bf 1302}, 085 (2013)
  [arXiv:1211.2472 [hep-ex]].

\bibitem{CMS:dis-dijet} 
  [CMS Collaboration],
  CMS-PAS-EXO-12-038.

  
\bibitem{Aad:2012bwa} 
  G.~Aad {\it et al.}  [ATLAS Collaboration],
  arXiv:1212.1272 [hep-ex].

\bibitem{CMS:tracking} 
  [CMS Collaboration],
  CMS-PAS-TRK-10-001.


\bibitem{Aad:2013txa} 
  G.~Aad {\it et al.}  [ATLAS Collaboration],
  arXiv:1305.2284 [hep-ex].
  
\bibitem{Meade:2010ji} 
  P.~Meade, M.~Reece and D.~Shih,
  JHEP {\bf 1010}, 067 (2010)
  [arXiv:1006.4575 [hep-ph]].



\bibitem{Baak:2012kk} 
  M.~Baak, M.~Goebel, J.~Haller, A.~Hoecker, D.~Kennedy, R.~Kogler, K.~Moenig and M.~Schott {\it et al.},
  arXiv:1209.2716 [hep-ph].
   
\bibitem{Batell:2012ca} 
  B.~Batell, S.~Gori and L.~-T.~Wang,
  JHEP {\bf 1301}, 139 (2013)
  [arXiv:1209.6382 [hep-ph]].

\bibitem{Maksymyk:1993zm} 
  I.~Maksymyk, C.~P.~Burgess and D.~London,
  Phys.\ Rev.\ D {\bf 50}, 529 (1994)
  [hep-ph/9306267].
   
\bibitem{Burgess:1993mg} 
  C.~P.~Burgess, S.~Godfrey, H.~Konig, D.~London and I.~Maksymyk,
  Phys.\ Lett.\ B {\bf 326}, 276 (1994)
  [hep-ph/9307337].
    
\bibitem{Martin:2004id} 
  S.~P.~Martin, K.~Tobe and J.~D.~Wells,
  Phys.\ Rev.\ D {\bf 71}, 073014 (2005)
  [hep-ph/0412424].


\bibitem{Kane:1992kq} 
  G.~L.~Kane, C.~F.~Kolda and J.~D.~Wells,
  Phys.\ Rev.\ Lett.\  {\bf 70}, 2686 (1993)
  [hep-ph/9210242].
  
\bibitem{Espinosa:1992hp} 
  J.~R.~Espinosa and M.~Quiros,
  Phys.\ Lett.\ B {\bf 302}, 51 (1993)
  [hep-ph/9212305].
  
\bibitem{Ellwanger:2009dp} 
  U.~Ellwanger, C.~Hugonie and A.~M.~Teixeira,
  Phys.\ Rept.\  {\bf 496}, 1 (2010)
  [arXiv:0910.1785 [hep-ph]].
  
\bibitem{Batra:2003nj} 
  P.~Batra, A.~Delgado, D.~E.~Kaplan and T.~M.~P.~Tait,
  JHEP {\bf 0402}, 043 (2004)
  [hep-ph/0309149].

\bibitem{Lu:2013cta} 
  X.~Lu, H.~Murayama, J.~T.~Ruderman and K.~Tobioka,
  arXiv:1308.0792 [hep-ph].
  
  
\bibitem{Huo:2012tw} 
  R.~Huo, G.~Lee, A.~M.~Thalapillil and C.~E.~M.~Wagner,
  Phys.\ Rev.\ D {\bf 87}, 055011 (2013)
  [arXiv:1212.0560 [hep-ph]].

\bibitem{Gherghetta:2012gb} 
  T.~Gherghetta, B.~von Harling, A.~D.~Medina and M.~A.~Schmidt,
  JHEP {\bf 1302}, 032 (2013)
  [JHEP {\bf 1302}, 032 (2013)]
  [arXiv:1212.5243 [hep-ph]].


\bibitem{Joglekar:2012vc} 
  A.~Joglekar, P.~Schwaller and C.~E.~M.~Wagner,
  JHEP {\bf 1212}, 064 (2012)
  [arXiv:1207.4235 [hep-ph]].

\bibitem{ArkaniHamed:2012kq} 
  N.~Arkani-Hamed, K.~Blum, R.~T.~D'Agnolo and J.~Fan,
  JHEP {\bf 1301}, 149 (2013)
  [arXiv:1207.4482 [hep-ph]].
  
\bibitem{Reece:2012gi} 
  M.~Reece,
  New J.\ Phys.\  {\bf 15}, 043003 (2013)
  [arXiv:1208.1765 [hep-ph]].
  
\bibitem{Batell:2012mj} 
  B.~Batell, D.~McKeen and M.~Pospelov,
  JHEP {\bf 1210}, 104 (2012)
  [arXiv:1207.6252 [hep-ph]].
  
\bibitem{Kearney:2012zi} 
  J.~Kearney, A.~Pierce and N.~Weiner,
  Phys.\ Rev.\ D {\bf 86}, 113005 (2012)
  [arXiv:1207.7062 [hep-ph]].
  
\bibitem{McKeen:2012av} 
  D.~McKeen, M.~Pospelov and A.~Ritz,
  Phys.\ Rev.\ D {\bf 86}, 113004 (2012)
  [arXiv:1208.4597 [hep-ph]].
  
\bibitem{Davoudiasl:2012ig} 
  H.~Davoudiasl, H.~-S.~Lee and W.~J.~Marciano,
  Phys.\ Rev.\ D {\bf 86}, 095009 (2012)
  [arXiv:1208.2973 [hep-ph]].
  
\bibitem{Batell:2012zw} 
  B.~Batell, S.~Jung and H.~M.~Lee,
  JHEP {\bf 1301}, 135 (2013)
  [arXiv:1211.2449 [hep-ph]].
  
\bibitem{Fan:2013qn} 
  J.~Fan and M.~Reece,
  JHEP {\bf 1306}, 004 (2013)
  [arXiv:1301.2597 [hep-ph]].
   
\bibitem{Joglekar:2013zya} 
  A.~Joglekar, P.~Schwaller and C.~E.~M.~Wagner,
  JHEP {\bf 1307}, 046 (2013)
  [arXiv:1303.2969 [hep-ph]].
  
\bibitem{Altmannshofer:2013zba} 
  W.~Altmannshofer, M.~Bauer and M.~Carena,
  arXiv:1308.1987 [hep-ph].
  


  
\end{thebibliography}
\end{document}